\documentclass[11pt]{article}
\pdfoutput=1

\usepackage{jinstpub}

\usepackage{paralist}
\usepackage{graphicx}
\usepackage{wrapfig}
\usepackage{lineno}
\usepackage{xcolor}
\usepackage{setspace}
\usepackage{hyperref}
\usepackage{rotating}

\graphicspath{{./Figures/}}

\title{The \textsc{Majorana Demonstrator} Readout Electronics System}

\newcommand{\ITEP}{National Research Center ``Kurchatov Institute'' Institute for Theoretical and 
Experimental Physics, Moscow, 117218 Russia}
\newcommand{\lbnl}{Nuclear Science Division, Lawrence Berkeley National Laboratory, Berkeley, CA 94720, USA}
\newcommand{\lanl}{Los Alamos National Laboratory, Los Alamos, NM 87545, USA}
\newcommand{\queens}{Department of Physics, Engineering Physics and Astronomy, Queen's University, Kingston, 
ON K7L 3N6, Canada}
\newcommand{\uw}{Center for Experimental Nuclear Physics and Astrophysics, and Department of Physics, 
University of Washington, Seattle, WA 98195, USA}

\newcommand{\unc}{Department of Physics and Astronomy, University of North Carolina, Chapel Hill, NC 27514, USA}
 
\newcommand{\ucph}{Department of Physics, University of California, Berkeley, CA 94720, USA}
\newcommand{\duke}{Department of Physics, Duke University, Durham, NC 27708, USA}
\newcommand{\ncsu}{Department of Physics, North Carolina State University, Raleigh, NC 27695, USA}	
\newcommand{\ornl}{Oak Ridge National Laboratory, Oak Ridge, TN 37830, USA}
\newcommand{\ou}{Research Center for Nuclear Physics, Osaka University, Ibaraki, Osaka 567-0047, Japan}
\newcommand{\pnnl}{Pacific Northwest National Laboratory, Richland, WA 99354, USA}
\newcommand{\williams}{Physics Department, Williams College, Williamstown, MA 01267, USA}
\newcommand{\ttu}{Tennessee Tech University, Cookeville, TN 38505, USA}
\newcommand{\sdsmt}{South Dakota School of Mines and Technology, Rapid City, SD 57701, USA}
\newcommand{\sjtu}{Shanghai Jiao Tong University, Shanghai, China}
\newcommand{\usc}{Department of Physics and Astronomy, University of South Carolina, Columbia, SC 29208, USA}
\newcommand{\usd}{Department of Physics, University of South Dakota, Vermillion, SD 57069, USA} 
\newcommand{\ut}{Department of Physics and Astronomy, University of Tennessee, Knoxville, TN 37916, USA}
\newcommand{\tunl}{Triangle Universities Nuclear Laboratory, Durham, NC 27708, USA}
\newcommand{\mpi}{Max-Planck-Institut f\"{u}r Physik, M\"{u}nchen, 80805 Germany}
\newcommand{\tum}{Physik Department and Excellence Cluster Universe, Technische Universit\"{a}t, M\"{u}nchen, 85748 Germany}
\newcommand{\lbleg}{Engineering Division, Lawrence Berkeley National Laboratory, Berkeley, CA 94720, USA}

\affiliation[a]{\lbnl}
\affiliation[b]{\pnnl}
\affiliation[c]{\usc}
\affiliation[d]{\ornl}
\affiliation[e]{\ITEP}
\affiliation[f]{\usd}
\affiliation[g]{\unc}
\affiliation[h]{\tunl}
\affiliation[i]{\sdsmt}
\affiliation[j]{\uw}
\affiliation[k]{\duke}
\affiliation[l]{\lanl}
\affiliation[m]{\ut}
\affiliation[n]{\ou}
\affiliation[o]{\williams}
\affiliation[p]{\ncsu}
\affiliation[q]{\ttu}
\affiliation[r]{\ucph}
\affiliation[s]{\sjtu}
\affiliation[t]{\lbleg}
\affiliation[u]{\queens} 
\affiliation[v]{\mpi}
\affiliation[w]{\tum}

\author[a]{N.~Abgrall,}	
\author[a]{M.~Amman,}
\author[b]{I.J.~Arnquist,} 
\author[c,d]{F.T.~Avignone~III,}
\author[e]{A.S.~Barabash,}
\author[f]{C.J.~Barton,}	
\author[a]{P.J.~Barton,}	
\author[d]{F.E.~Bertrand,}
\author[g,h]{K.H.~Bhimani,}
\author[i,g,h]{B.~Bos,}		
\author[a,1]{A.W.~Bradley,~\note{Present address: 775 Heinz Ave, Berkeley, CA 94710, USA}}	
\author[j]{T.H.~Burritt,}
\author[k,h]{M.~Busch,}
\author[j,2]{M.~Buuck,~\note{Present address: SLAC National Accelerator Laboratory, Menlo Park, 
CA 94025, USA}} 
\author[g,h]{T.S.~Caldwell,}	
\author[a]{Y-D.~Chan,}
\author[i]{C.D.~Christofferson,}
\author[l]{P.-H.~Chu,}
\author[g,h]{M.L.~Clark,}
\author[d,3]{R.J.~Cooper,~\note{Present address: \lbnl}} 
\author[j,4]{C. Cuesta,\note{Present address: Centro de Investigaciones Energ\'eticas, Medioambientales 
y Tecnol\'ogicas, CIEMAT, 28040, Madrid, Spain}}
\author[j]{J.A.~Detwiler,}	
\author[a]{A.~Drobizhev,}
\author[c]{D.W.~Edwins,}
\author[m,d]{Yu.~Efremenko,}
\author[n]{H.~Ejiri,}
\author[l]{S.R.~Elliott,}
\author[g,h]{T.~Gilliss,}
\author[o]{G.K.~Giovanetti,}  
\author[p,h,d]{M.P.~Green,} 
\author[g,h]{J.~Gruszko,}	
\author[g,h]{I.S.~Guinn,}		
\author[d]{V.E.~Guiseppe,}	
\author[g,h]{C.R.~Haufe,}
\author[g,h]{R.J.~Hegedus,}
\author[g,h]{R.~Henning,}
\author[g,h]{D.~Hervas~Aguilar,}
\author[b]{E.W.~Hoppe,}
\author[j]{A.~Hostiuc,}
\author[q]{M.F.~Kidd,}
\author[l]{I.~Kim,}
\author[b]{R.T.~Kouzes,}
\author[g,h]{A.~Li,}
\author[s,a]{J.C.~Loach,}	
\author[m]{A.M.~Lopez,}
\author[f]{J.M.~L\'opez-Casta\~no,}
\author[t,a]{P.N.~Luke,}
\author[g,h]{E.L.~Martin,}
\author[u]{R.D.~Martin,}	
\author[l]{R.~Massarczyk,}		
\author[l]{S.J.~Meijer,}
\author[v,w]{S.~Mertens,}		
\author[a]{J.~Myslik,}	
\author[f]{T.K.~Oli,}
\author[g,h]{G.~Othman,} 
\author[j]{D.~Peterson,}
\author[j,5]{W.~Pettus,~\note{Present address: Center for Exploration of Energy and Matter, Indiana University, 
Bloomington, IN 47408, USA}}	
\author[a,6]{A.W.P.~Poon,~\note{Corresponding author.}}
\author[d]{D.C.~Radford,}
\author[g,h]{J.~Rager,}
\author[g,h]{A.L.~Reine,}	
\author[l]{K.~Rielage,}
\author[j]{R.G.H.~Robertson,}
\author[j]{N.W.~Ruof,}
\author[l]{B.~Sayki,}
\author[l]{M.J.~Stortini,}
\author[c]{D.~Tedeschi,}
\author[a]{M.~Turqueti,}
\author[j]{T.D.~Van Wechel,}
\author[d]{R.L.~Varner,}
\author[w]{S.~Vasilyev,}
\author[a,7]{K.~Vetter,~\note{Alternate address: Department of Nuclear Engineering, University of California, 
Berkeley, CA, USA}}
\author[h,i,d]{J.F.~Wilkerson,}    
\author[j]{C.~Wiseman,}		
\author[f]{W.~Xu,}
\author[t]{H.~Yaver,}
\author[d]{C.-H.~Yu,}
\author[l,8]{B.X.~Zhu,~\note{Present address: Jet Propulsion Laboratory, California Institute of Technology, 
Pasadena, CA 91109, USA}} 
\author[t]{and S.~Zimmermann} 

\emailAdd{awpoon@lbl.gov}


\collaboration[c]{(\textsc{Majorana} Collaboration)}

\abstract{The \textsc{Majorana Demonstrator} comprises two arrays of high-purity
    germanium detectors constructed
    to search for neutrinoless double-beta decay in $^{76}$Ge
    and other physics beyond the Standard Model. Its
    readout electronics were designed to have low electronic noise, and
    radioactive backgrounds were minimized by using low-mass components and
    low-radioactivity materials near the detectors. 
    This paper provides a description of all components 
    of the \textsc{Majorana Demonstrator} readout electronics, spanning the
    front-end electronics and internal cabling, back-end electronics,
    digitizer, and power supplies, along with the grounding scheme.  The
    spectroscopic performance achieved with these readout electronics is also
    demonstrated.} 
\begin{document}

\maketitle
\flushbottom

\section{Introduction}
\label{sec_introduction}

The {\sc Majorana Demonstrator}'s~\cite{MAJORANA} searches for neutrinoless
double-beta decay~\cite{Aalseth:2017btx,Alvis:2019sil,Dolinski:2019nrj} and other physics 
beyond the  Standard Model place stringent
requirements on the design of its readout electronics.  The need to separate
neutrinoless double-beta decay signal events at 2.039~MeV (the Q-value of $^{76}$Ge
double-beta decay) from the two neutrino double-beta decay events (a continuum
with energy up to the Q-value) necessitates excellent energy resolution, and
therefore low-noise readout electronics.  To realize a sensitive search, backgrounds induced by 
cosmic rays and other natural radioactivity are minimized
by locating the experiment underground at the 4850-ft level of the Sanford
Underground Research Facility (SURF)~\cite{Heise:2017rpu} in Lead,~SD,~USA, and by judicious selection of detector
construction materials~\cite{Abgrall:2016cct}. These features also enable
sensitive searches for physics beyond the Standard Model at low energies
($<100$~keV)~\cite{Abgrall:2016tnn,Alvis:2018yte}.

\begin{figure}[!htb]
 \includegraphics[width=\columnwidth]{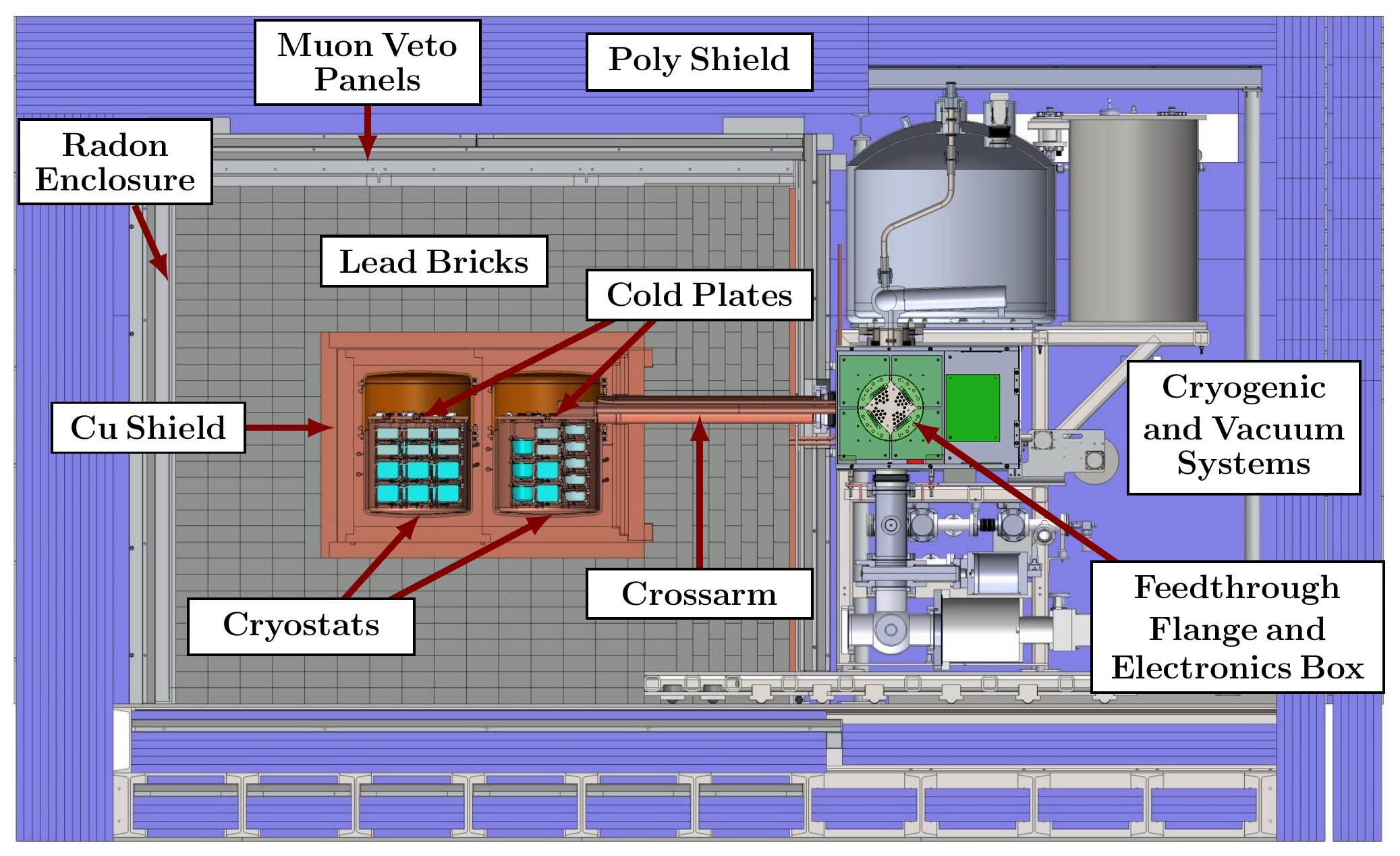}
 \caption{Schematic diagram (cutaway view) of the \textsc{Majorana Demonstrator}.  The
  cryogenic and vacuum systems are shown only for the cryostat on the right.
  The second feedthrough flange and electronics box for this cryostat are on the opposite side of the electronics box shown.
 \label{fig:MJD-diagram}}
\end{figure}

The {\sc Majorana Demonstrator} (Figure~\ref{fig:MJD-diagram}) comprises two independent modules,
each containing a vacuum cryostat with its own separate vacuum and cryogenic
systems. Inside each cryostat, seven~\emph{strings} holding up to five p-type high-purity
germanium (HPGe) detectors each are suspended from the copper \emph{cold plate}, a large
conductive thermal mass to conduct heat away from the detectors.  The cryostats
are surrounded by copper and lead shielding, enclosed in a radon exclusion
box.  Both the inner shield and the copper parts inside the cryostats are made out of 
electroformed copper produced and machined in the \textsc{Majorana Demonstrator}'s underground
laboratory \cite{osti_1039850, Christofferson:2017nih}.  The active muon veto
system~\cite{Abgrall:2016cfi, MAJORANA}, which comprises plastic scintillator panels, locates outside the radon enclosure.
Each cryostat is connected through the shielding to its vacuum and cryogenic systems and its
warm electronics via the \emph{crossarm}. Two custom vacuum feedthrough flanges
are located at the end of each crossarm. Mounted directly
on each flange are the electronics boxes that contain the back-end
electronics. Polyethylene shielding (``poly shield'') surrounds all of this apparatus.  The
digitizers and power supplies are located in equipment racks outside the poly
shield.  

Each of the two modules consists of both $^{76}$Ge-enriched p-type point-contact (PPC) detectors~\cite{LukePPC, Barbeau:2007qi} and commercial Broad Energy Germanium (BEGe) detectors~\cite{CANBERRA}.  The bulk semiconducting material of the PPC detectors is p-type.  The  
small single ``point'' contact for charge collection in a PPC detector results in a small capacitance of $\sim$2~pF.  
Each detector's bias voltage is applied on the n+ contact. 
The p+ contact is electrically connected to a low-mass 
front-end (LMFE) board~\cite{LMFE} for signal charge amplification. The
internal cabling routes from the detectors to the cold plate, then through
the crossarm to the feedthrough flanges and the electronics boxes.
The mapping of detectors to the power supplies and components of the readout 
electronics is designed to be flexible.  In the configuration implemented in the \textsc{Majorana
Demonstrator}, the detectors of a string share a common high-voltage power-supply module.  Individual channels of the module can supply each detector with a different operation voltage (HV,
0 to 5000~V).  Each
custom-built low-voltage (LV, -24 to 24~V) power supply powers
the motherboards, preamplifier cards, front-ends, and a
controller card for two strings of detectors. The
differential output signals from the preamplifier cards are
sent to digitizer cards. 

From May to October~2015, only Module~1 was instrumented.  It contained twenty $^{76}$Ge-enriched PPC detectors with a total mass of 16.8~kg along with nine natural BEGe detectors with a total mass of 5.6~kg.  Module 2 was deployed in mid-2016 with fifteen $^{76}$Ge-enriched PPC detectors with a total mass of 12.9~kg and fourteen natural BEGe detectors with a total mass of 8.8~kg.  

\begin{figure}[!htpb]
 \includegraphics[width=0.97\linewidth]{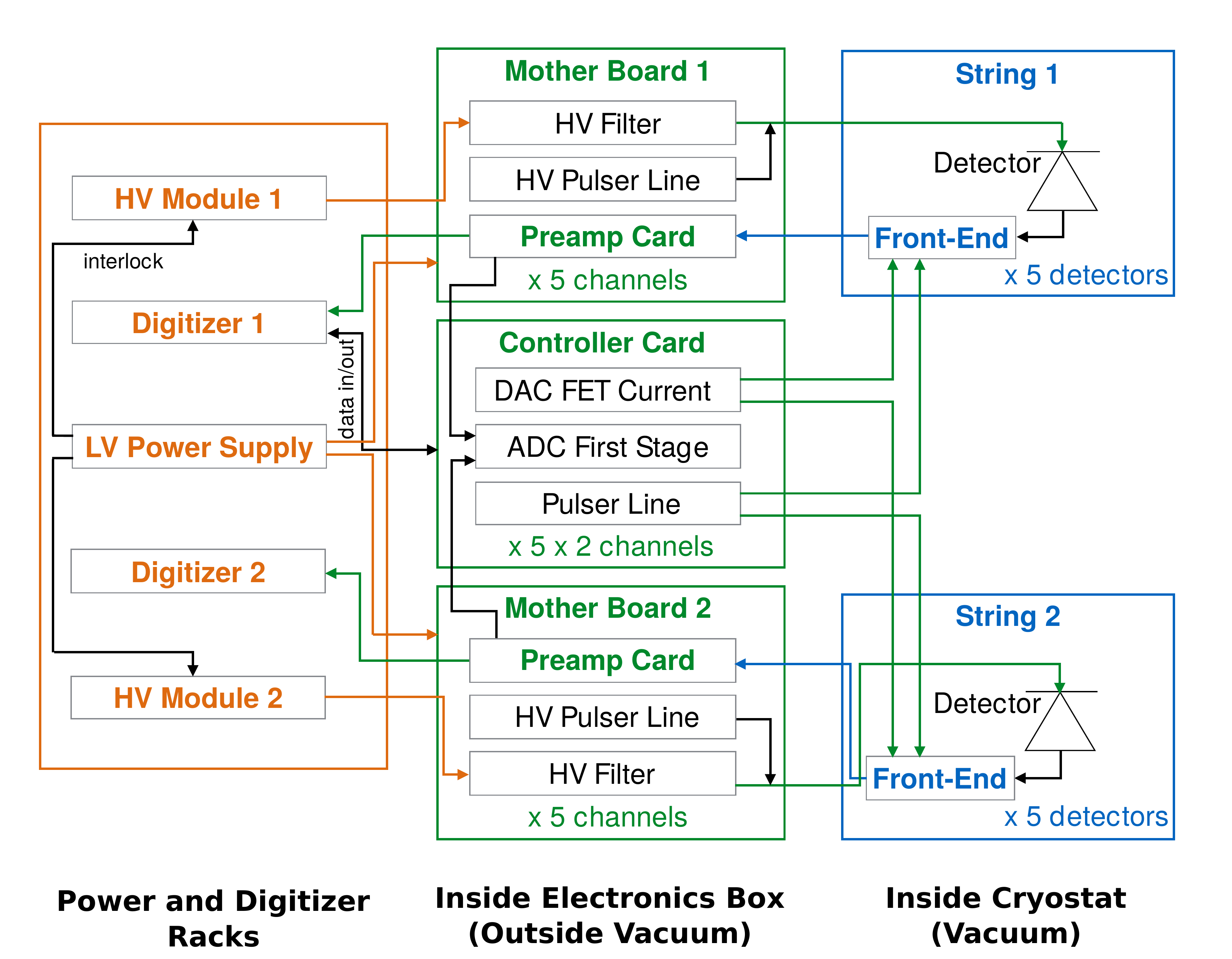}
 \caption{The \textsc{Majorana Demonstrator} readout electronics configuration for 2 strings of
up to 5 detectors, each consisting of the power supplies and digitizers (left), the back-end
(middle) and the front-end electronics (right).  Power to
the controller cards, preamplifier cards, and front-ends is provided through the
motherboard's connections to the low-voltage power supply.  In this figure, we have simplified the connections between individual channels for improved readability.  For example, the output from a high-voltage power-supply module is represented by an arrow, although each output channel of these  modules is typically mapped to a detector in a string.  Thus, an external pulser can inject charges through the high-voltage line to a detector and calibrate its low-energy response.  A separate on-board pulser in the controller board can also inject charges to a detector's front-end for live time measurement.
 \label{fig_electronics_general_layout}}
\end{figure}

The default electronics
configuration is shown in Figure~\ref{fig_electronics_general_layout}.
The order of presentation of this paper follows the path of the charges
collected from the detectors, approximately from right to left in the figure.  In Sections~\ref{sec_front_end_electronics} and \ref{sec_internal_cabling},
we describe the front-end electronics and the internal cabling inside the vacuum cryostats. 
Section~\ref{sec_back_end_electronics} presents the back-end
electronics outside the vacuum. In Section~\ref{sec_power_supplies_and_digitizer}, we describe the signal digitizer card
along with the high-voltage and low-voltage
 power supplies.  In 
Section~\ref{sec_general_grounding_scheme}, we review the grounding scheme of
the experiment. The spectroscopic performance of the \textsc{Majorana
Demonstrator} is shown last in Section~\ref{sec:spec_perf}.

\section{Front-end electronics}
\label{sec_front_end_electronics}

The low-mass front-end~\cite{LMFE} board is the first element in 
the detector signal amplification chain.  They are mounted in close proximity
($\sim 1$~cm) to each detector's point contact to minimize noise. 
To enable their placement so close to the detectors, radioactive backgrounds in the
LMFE were minimized through its low-mass design and careful selection of
components~\cite{Abgrall:2016cct}. 

\begin{figure}[!htb]
\begin{centering}
    \hspace{0.5cm}
    \includegraphics[width=0.45\linewidth]{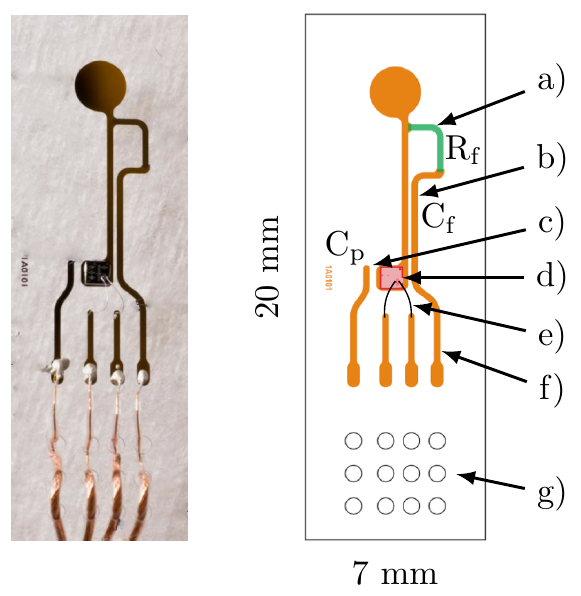}
    \caption{A photo (left) and a labeled diagram (right) of an
    LMFE.  The circular pad at the top connects to the contact pin, and the
    four traces at the bottom  are from left to right: pulser, drain,
    source, feedback.  The labels are  a) feedback resistor (R$_{\mathrm{f}}$);
    b) feedback capacitance (C$_{\mathrm{f}}$) and c) pulser capacitance
    (C$_{\mathrm{p}}$) are both provided by the capacitance between traces; d) MX-11 
    JFET bare die;
    e) wire bonds; f) Ti/Au traces; and g) strain relief holes in fused silica
    substrate.}
    \label{fig_lmfe_diagram}
    \hspace{0.5cm}
\end{centering}
\end{figure}

The LMFE (Figure~\ref{fig_lmfe_diagram}) is built upon a rectangular 20~mm~x~7~mm substrate of 200~$\mu$m thick
fused silica (Corning\textsuperscript{\textregistered}~7980).  On each LMFE is a
bare-die Moxtek MX-11~\cite{MOXTEK} junction gate field-effect transistor
(JFET) whose properties are shown in
Table~\ref{tab:mx11prop}, a feedback resistor ($\sim10$~G$\Omega$), and a
feedback capacitor (0.17~pF) in a resistive-feedback configuration.  The feedback capacitance is provided
by the capacitance between traces, as is the 45~fF capacitance used for
injecting pulser signals (discussed in
Section~\ref{subsec_CC}) to the gate of the JFET.  The traces are sputtered 20/400~nm Ti/Au, and the
drain and source of the JFET are wire-bonded to the traces with
25.4~$\mu$m Al(1\%~Si) wires.  The gate on the bottom of the JFET is affixed 
to its trace on the fused silica with low-outgassing silver epoxy (Henkel Hysol\textsuperscript{\textregistered} 
TRA-DUCT 2902).  This epoxy is stable under temperature cycling and 
has demonstrated acceptable radiopurity under assays.
\begin{table}
    \caption{Electronic properties of MX-11 JFETs, provided by
    Moxtek~\cite{MOXTEK}. V$_{\mathrm{DS}}$ denotes the drain-source voltage,
    I$_{\mathrm{D}}$ denotes the drain current, and V$_{\mathrm{GS}}$ denotes
    the gate-source voltage.  Many properties are highly temperature-dependent
    (especially the leakage current and transconductance). NB: Other MX-11 variants
    (e.g.~MX-11rc) have different properties.\label{tab:mx11prop}}

    \centering
    \begin{tabular}{|c|c|c|}
        \hline
    Property & Measurement Conditions & Value \\ \hline
        Leakage Current & Single Junction, T~$=20^{\circ}$C & $<1$~pA\\ \hline
        Transconductance (g$_{\mathrm{m}}$) &V$_{\mathrm{DS}}=4$~V,
        I$_{\mathrm{D}}=5$~mA, T~$=20^{\circ}$C & 5.6 mS\\
         &V$_{\mathrm{DS}}=2$~V,
        I$_{\mathrm{D}}=1.5$~mA, T~$=20^{\circ}$C & 4.1 mS\\ \hline
        Cut-off Voltage (V$_{\mathrm{off}}$) & V$_{\mathrm{DS}}=2$~V,
        I$_{\mathrm{D}}=1$~nA & --4.2~V\\ \hline
        Drain Saturation Current (I$_{\mathrm{DSS}}$) &V$_{\mathrm{DS}}=4$~V,
        V$_{\mathrm{GS}}=0$~V & 30~mA\\ \hline
        Gate-to-Source Capacitance (C$_{\mathrm{GS}}$) & V$_{\mathrm{DS}}=2$~V,
        I$_{\mathrm{D}}=1.5$~mA, 1~MHz & 2.70~pF\\ \hline

    \end{tabular}
\end{table}

\begin{figure}[!htb]
\begin{centering}
    \includegraphics[width=0.45\linewidth]{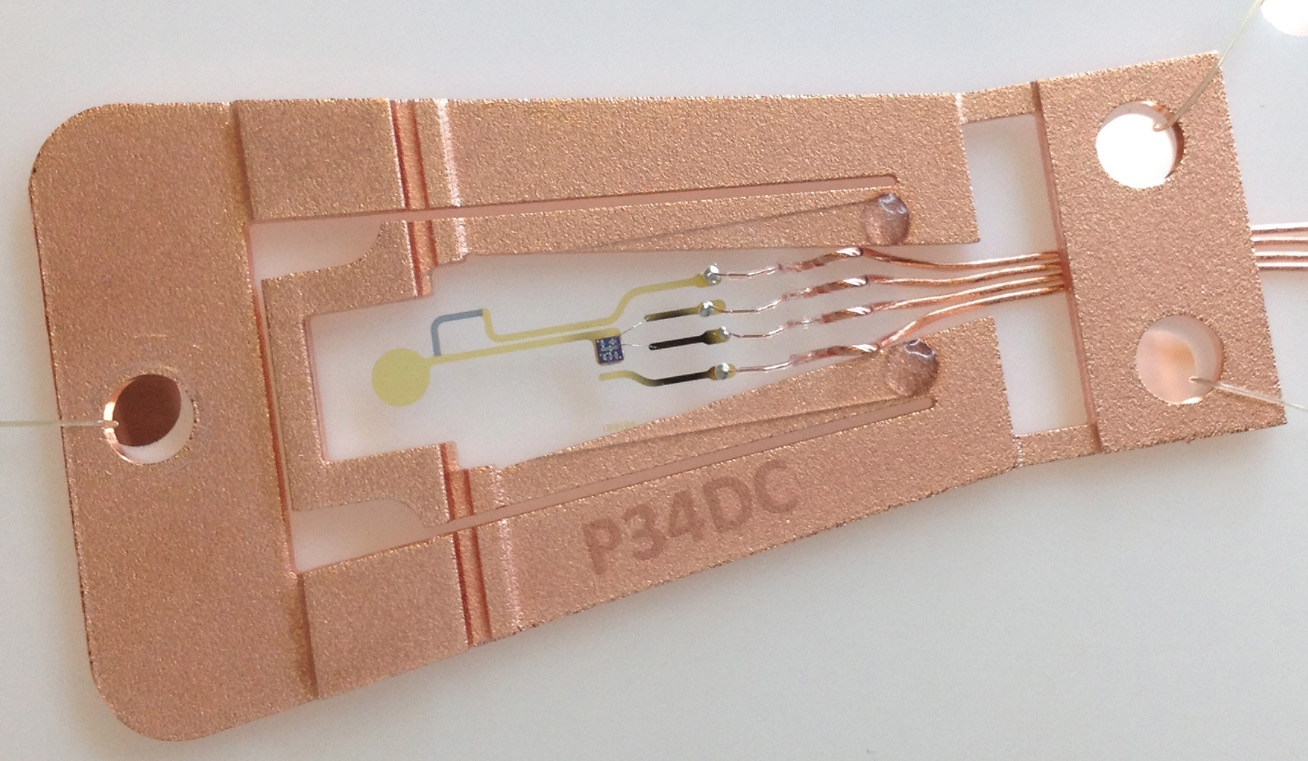}
    \includegraphics[width=0.45\linewidth, trim=0 0 0 4.5cm, clip]{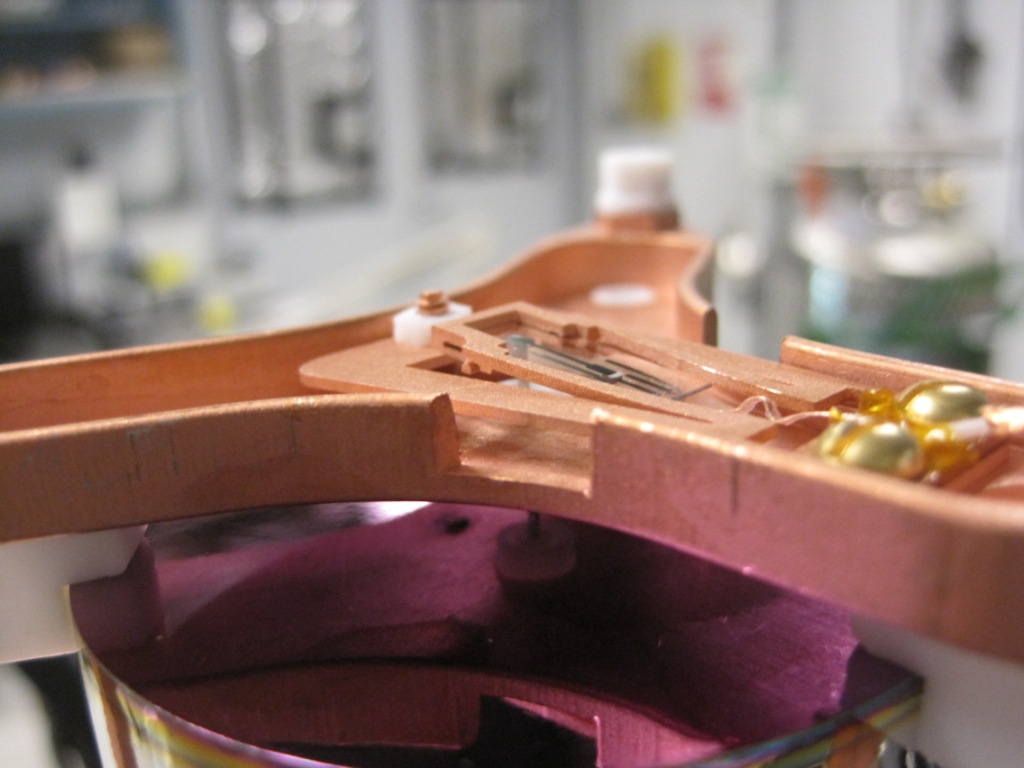}
    \includegraphics[width=0.45\linewidth]{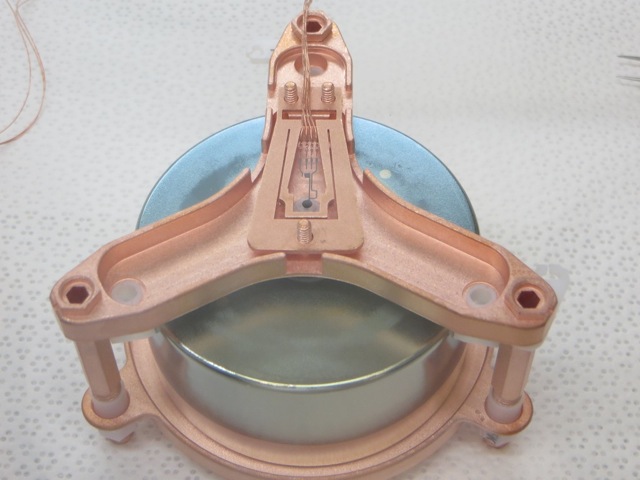}
    \includegraphics[width=0.45\linewidth,trim=0 4.1cm 0 60cm, clip]{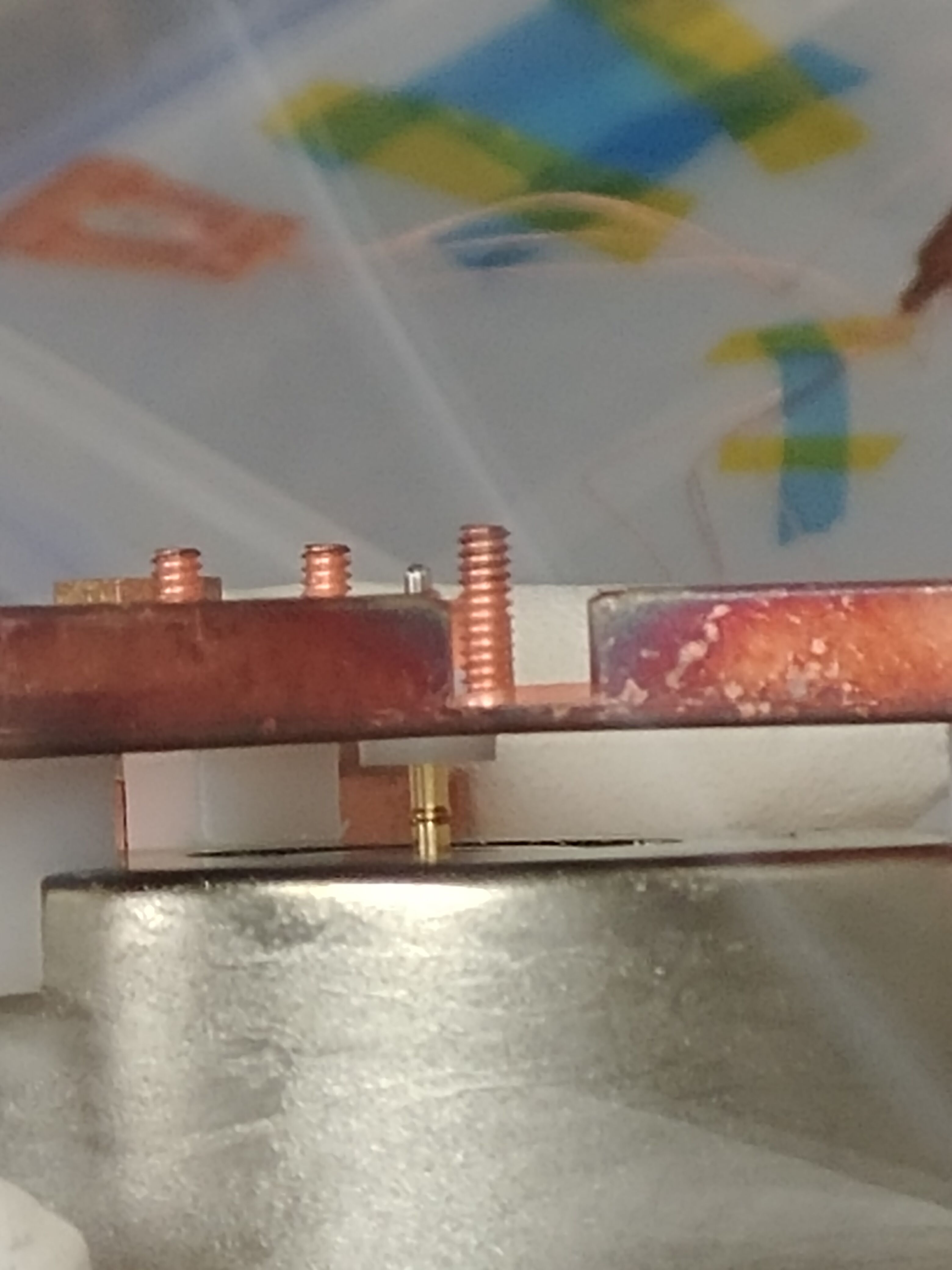}
    
    \caption{(Top left) An LMFE mounted in a spring clip.  The spring clip's
    \textsc{Majorana} Parts Tracking
    Database~\cite{Abgrall:2015jfa} serial number (P34DC) is laser-engraved. (Top right)
    An LMFE mounted in a spring clip above a detector during a
    bench test (strain relief holes were not present, temporary brass screws and Kapton tape were used, and 
	PTFE nuts were used instead of parylene-coated copper nuts).  The spring clip is
    under tension in this photo, demonstrating its spring action.
    (Bottom left) An LMFE in a spring clip mounted on a detector (before securing 
    parylene-coated copper nuts).
    (Bottom right) Side view of the contact pin pushing
    up from the detector through the Teflon bushing, prior to LMFE
    installation.}
    \label{fig_lmfe_spring_clip}
    \hspace{0.5cm}
\end{centering}
\end{figure}

Each LMFE is mounted in a custom-designed spring clip,
machined in an underground cleanroom out of electroformed copper.  The circular
pad end of the LMFE is inserted into a slot in the clip, and the other end is fixed onto the clip with Henkel~Hysol~0151 epoxy.  Photographs
of the LMFE in a spring clip and the surrounding detector-mount components are shown in
Figure~\ref{fig_lmfe_spring_clip}. The electrical connection between the LMFE readout pad and the
detector's charge-collecting electrode is made by a contact pin $1$~cm long and $0.1$~cm in diameter.  The pin is made out
of electroformed copper and coated at both ends with high-purity tin provided
by Canberra (now Mirion Technologies (Canberra) Inc~\cite{CANBERRA}).  The spring-clip aligns the
LMFE pad with the contact pin, which is held in place by a
polytetrafluoroethylene (PTFE) bushing.  
Three \#4-40 electroformed copper nuts, which have been coated with parylene as a lubricant to prevent galling, are used to secure the spring clip to the
detector mount.  The end of the spring clip with the cable connections is
attached to the detector mount with two of these nuts, while the other end is fastened to a threaded
stud with one nut. Turning this nut controls the tension on the contact pin. 
 In the \textsc{Majorana Demonstator}, this nut is tightened to provide just
 under 7~N of force between the LMFE and the contact pin, which ensures good
 electrical contact and reduces microphonic noise without damaging the fused silica substrate.

\begin{figure}[!htb]
\centering
    \includegraphics[width=\columnwidth]{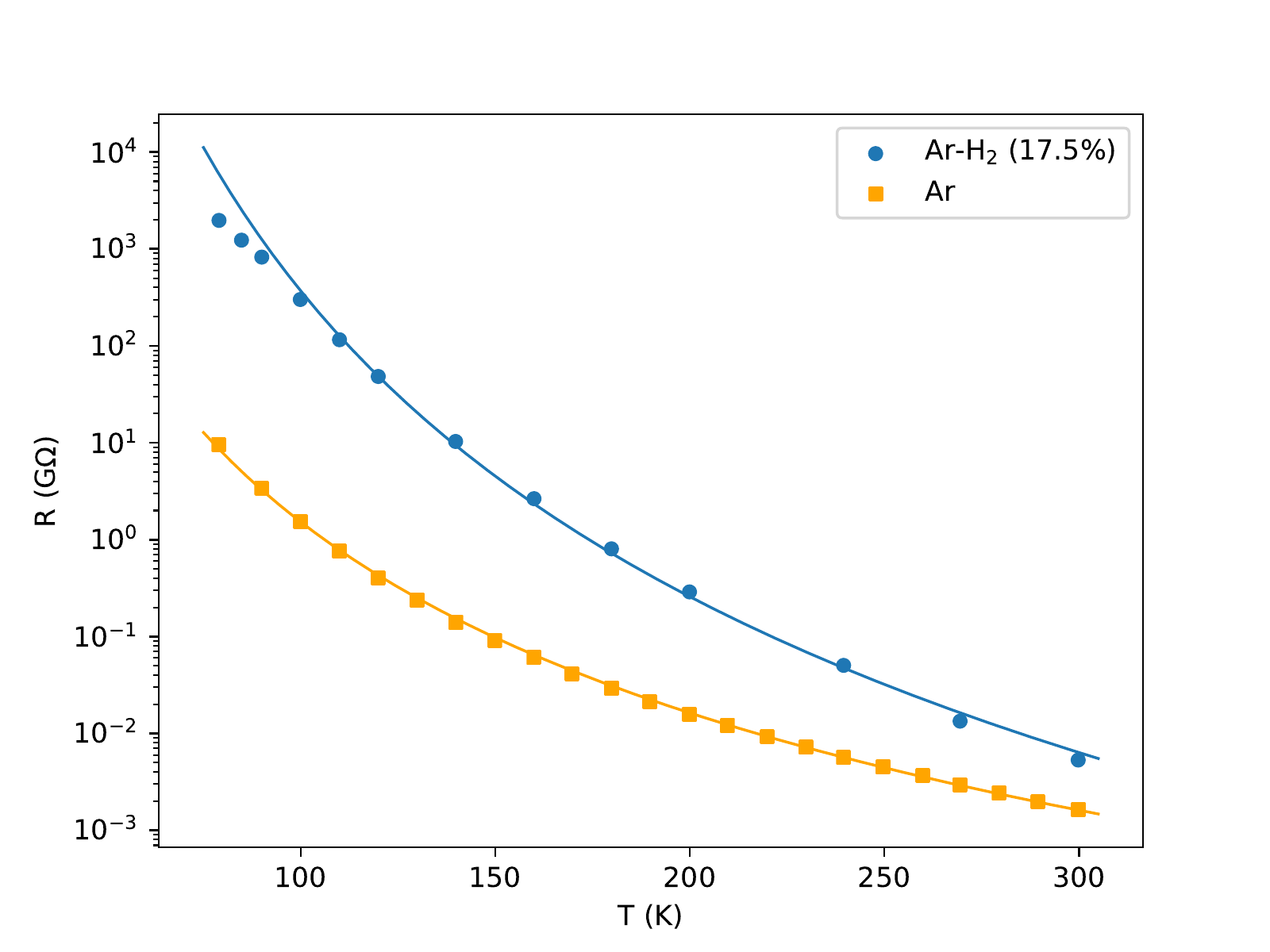}
    \caption{Resistance vs. temperature measurements for amorphous germanium resistors
    sputtered in different carrier gases (Ar, and Ar-H$_{2}$
    (17.5\%)).  The best-fit to $\mathrm{R}=\mathrm{e}^{\mathrm{a} +
    \mathrm{bT}^{-1/4}}$  is shown for both gases. Argon carrier gas was used to prepare the resistors on the LMFE
    boards due to its lower resistance at detector operating temperature. \label{fig:ge-resistance-vs-t}
    }

\end{figure}

The feedback resistor is a 400-nm-thick film of sputtered amorphous germanium,
which has low 1/f noise and radioactive background.  The resistance of the film varies with temperature T (in Kelvin) as exp(T$^{-\frac{1}{4}}$)
(Figure~\ref{fig:ge-resistance-vs-t}), requiring a stable operating temperature.  Any resulting variances in the preamplifier decay times between detector channels can be monitored with in-situ pulsing and are readily addressed during pulse shape analysis of the digitized pulse data~\cite{Alvis:2019sil}.

\begin{figure}[!htb]
\begin{centering}
    \hspace{0.5cm}
    \includegraphics[width=0.97\linewidth]{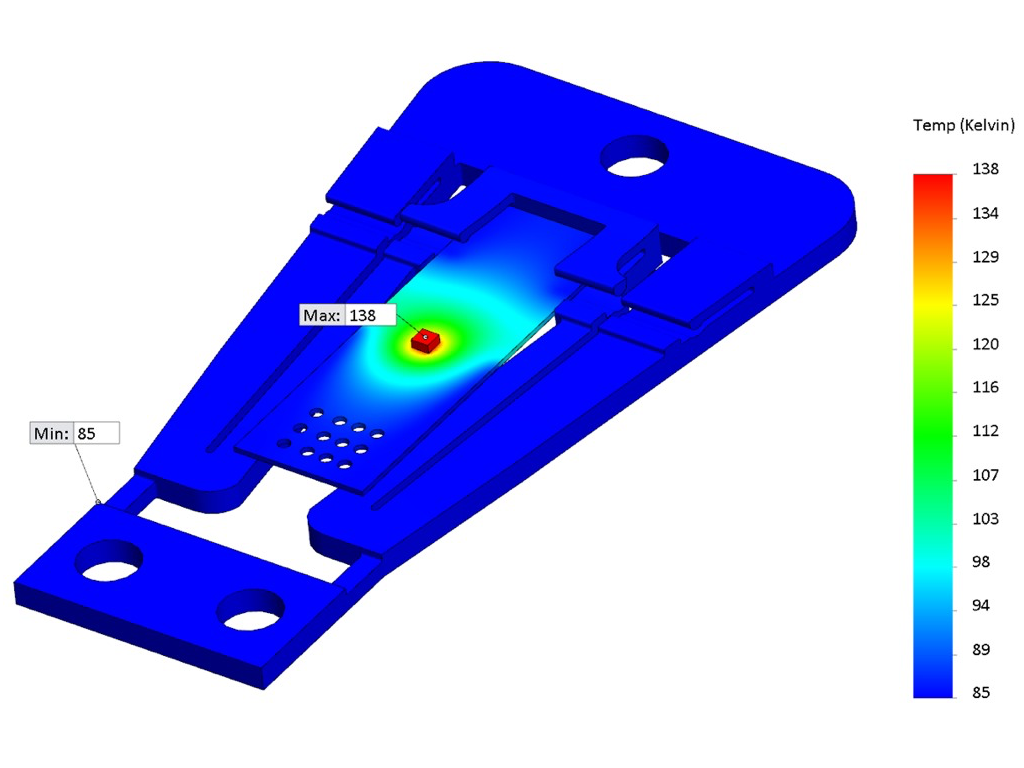}
    \caption{Thermal model of an LMFE mounted in an electroformed
    copper spring clip under normal operating conditions in the vacuum
    cryostat.  The spring-clip is at 85~K (as opposed to the liquid nitrogen
    temperature of 77~K) due to its mounting position in the detector string
    and heat loss within the detector system.}
    \label{fig_lmfe_temperature_gradient}
    \hspace{0.5cm}
\end{centering}
\end{figure}

The resistor geometry was
designed by considering the heat dissipation of the JFET and the resulting
temperature gradient on the LMFE to achieve the desired resistance (Figure~\ref{fig_lmfe_temperature_gradient}).
The drain-source voltage
(V$_{\mathrm{DS}}$) is user-adjustable between 0~V and 4.1~V through the controller card to compensate for temperature variances.  This adjustment can be made via the
ORCA (``Object-oriented Real-time Control and Acquisition'') system~\cite{ORCA}. A nominal value of $3.5$~V is
set to produce a $10$-mA drain current ($\mathrm{I}_{\mathrm{D}}$), resulting
in the JFET dissipating $35$~mW as heat. 
The temperatures of components on the
LMFE depend on the temperature of the copper clip, which is estimated to be at
approximately 85~K (as opposed to the liquid nitrogen temperature of 77~K) due
to its mounting position in the detector string and heat loss in the
detector system.  The operating temperature of the JFET
is about $130$~K, where near-optimal noise performance is achieved.  The temperature
gradient along the resistor results in its effective temperature at 
$90$~K, leading to the $10$~G$\Omega$ resistance
that optimizes noise performance.  This gives a decay time constant of
1.7~ms at the output of the first stage of preamplification. Combined with 
the capacitive coupling to the second stage of amplification (required since the first-stage baseline was outside the dynamic range of the digitizer), this results in a decay time constant of $\sim70\;\mu$s for the digitized signal.  This is suitable for both
the normal detector event rate ($\sim0.1-2$~Hz) and the higher event rate during
calibration ($\sim5-20$~Hz)~\cite{Abgrall:2017gpr}.

The LMFE has four traces for connections to an external pulser, JFET drain, JFET source, and the 
feedback capacitance and resistor in parallel.
These traces are each connected to the central conductor of an Axon' pico-coax
cable with the same silver epoxy used to attach the JFET gate. The cables are
woven through three rows of holes for strain relief.  These holes were ultrasonically drilled in
the substrate by Bullen~\cite{BULLEN}.

\begin{figure}
    \centering
    \includegraphics[width=0.49\columnwidth]{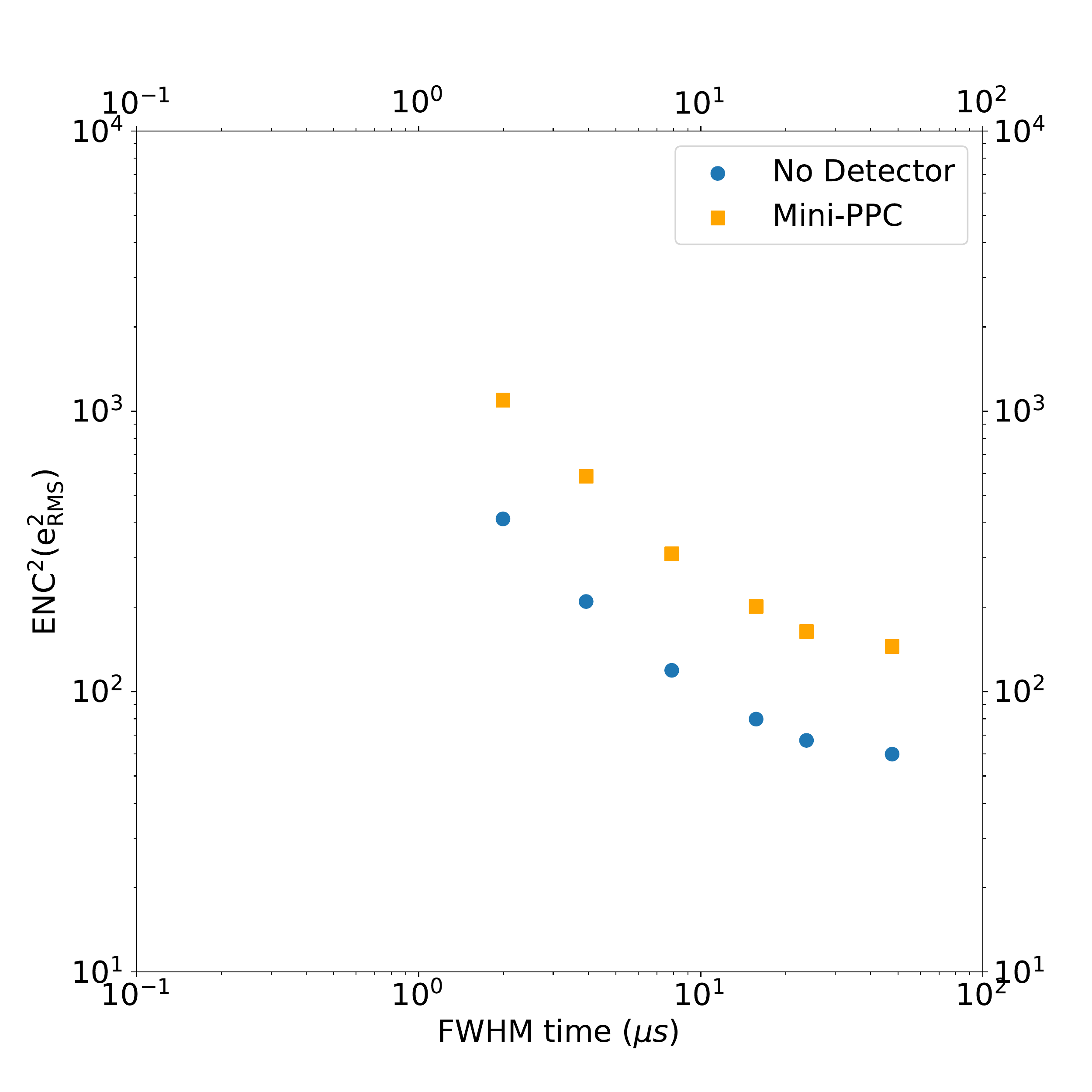}
    \includegraphics[width=0.49\columnwidth]{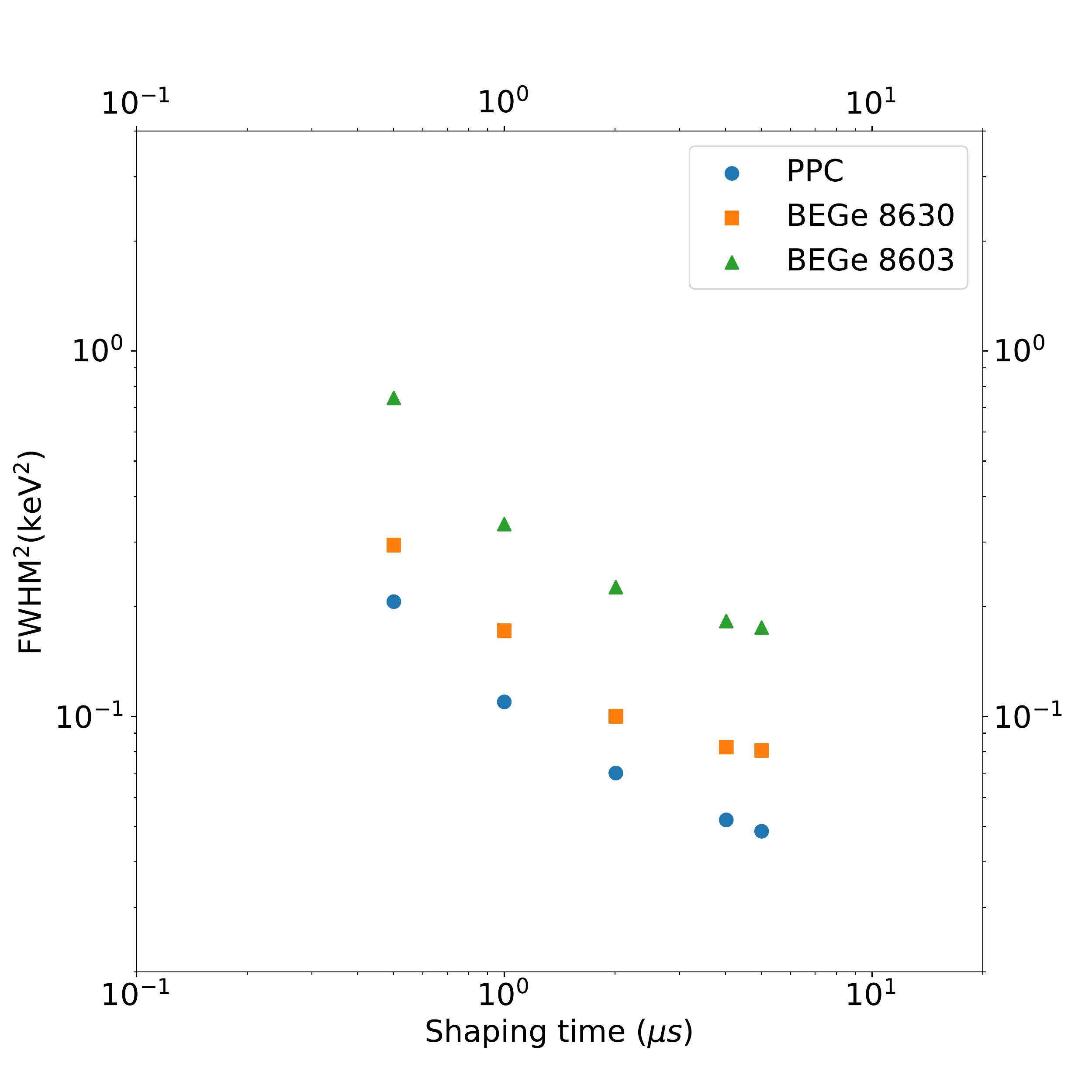}
    \caption{Noise measurements of an LMFE. A stable pulser was used to inject a fixed amount of charge through the detector into the LMFE. A radiation check source was then used to calibrate the amount of pulser-injected charge to a known energy.  (Left) Equivalent noise charge (squared) as a function of the FWHM equivalent shaping time.  Measurements for no additional capacitive load and with the mini-PPC mounted are shown.  The mini-PPC was operating at $\sim$82~K. (Right) Three-detector string mounted with prototype copper mounts in a test cryostat
    using the full production electronics readout chain.  Detectors in this string were operating in the temperature range of $\sim$85-88~K.}
    \label{fig_lmfe_string_noise_curve}
\end{figure}

Figure~\ref{fig_lmfe_string_noise_curve} shows the typical noise curves measured using an
LMFE-based readout under different conditions.  All detectors had low capacitance, approximately 2-3 pF.  As a result, the series noise was very low.  A prototype LMFE operated without the additional capacitive load of a detector
achieved a minimum noise level of 55~eV FWHM (full-width at half-maximum).  When mounted on a small
prototype PPC detector with a small p+ contact (the
``mini-PPC''~\cite{LMFE}), a minimum noise of 85~eV FWHM was achieved.
When mounted in a test string of three detectors (one \textsc{Majorana} PPC and two BEGe detectors) using the full \textsc{Majorana}
production electronics, a minimum noise level of 221~eV FWHM was achieved
with a integration time of 5.9~$\mu$s.

\section{Internal cabling}
\label{sec_internal_cabling}
\begin{figure}
    \centering
    \includegraphics[width=0.50\textwidth]{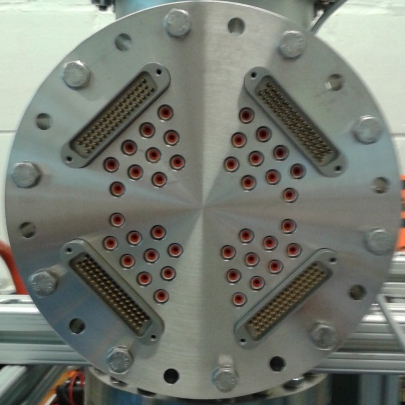}
    \includegraphics[trim=0 0 15cm 0,clip,width=0.297\textwidth]{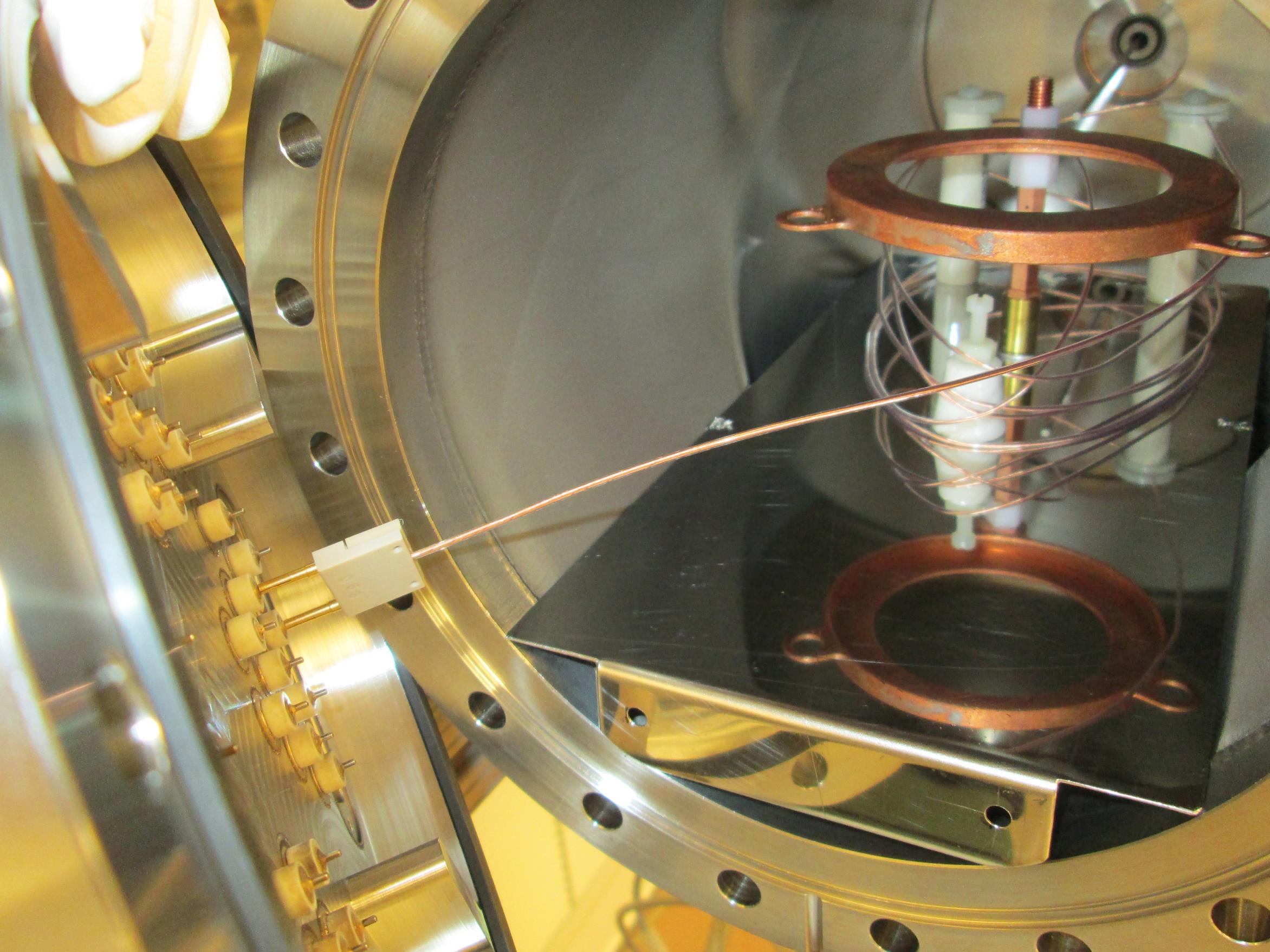}
    \caption{(Left) Picture of the feedthrough flange exterior.  The female end of the 40 Pee-Wee connectors 
    and the four 50-pin D-sub connectors are pictured.
    (Right) Side view of the vacuum-side of the
    feedthrough flange during a bench test.  The male end of the Pee-Wee
    connectors and an HV cable connected to two Pee-Wee connectors (for bias voltage and
    ground return) are shown.\label{fig:ftflange}}
\end{figure}

The internal cabling consists of two different coaxial cable designs.  The
\emph{signal cables} connect the pulser, drain, source, and 
feedback traces on each LMFE to the preamplifier card. 
The \emph{HV cables} deliver the bias voltage to the detectors.
The internal cabling for each module routes through two CF8 feedthrough flanges located outside the lead
shielding, as shown in Figure~\ref{fig:MJD-diagram}.
Each feedthrough flange has four 50-pin D-Sub connectors (DD-50).  Each
connector can connect to six detectors (a pin for each of the four signal cables and
their ground shields, for a total of 48 pins).  Each flange also has 40 ``Pee-Wee'' connectors (SRIPW101
from SRI~Hermetics), rated to 12~kV~DC, for the bias high voltage and returning ground for up to 20 detectors.
The feedthrough flange is shown in Figure~\ref{fig:ftflange}.

\subsection{Internal signal cabling}
Internal signal cables were custom produced by Axon' Cable in Montmirail,
France, using materials pre-screened for radiopurity. Based on the Axon'
PicoCoax\textsuperscript{\textregistered} design, the internal signal cables
consist of a 0.076-mm diameter central
conductor provided by California Fine Wire, an uncolored 0.254-mm outer diameter fluorinated ethylene propylene (FEP) dielectric, $28$ strands of 0.02504-mm diameter copper wires forming a helical
shield, and a 0.4-mm outer diameter FEP jacket.  The resulting cables have a characteristic impedance
of 55-58~$\Omega$ and a capacitance of 87~pF/m, depending on the production
batch. The linear mass density is 0.4~g/m. 

The total length of signal cable between each LMFE and the feedthrough
flange is 2.15~m.  This is broken up into two runs: one between the
LMFE and the cold plate, and the other between the cold plate and
the flange.  Four coaxial cables in a bundle are used for each LMFE, for each of the two runs. On the vacuum side of the flange, 
the signal cables are soldered to D-Sub sockets using low-radioactivity solder (one for the central
conductor, and one for the shield), which are then inserted to
the D-Sub receptacles on the flange.

\begin{figure}[!htb]
    \includegraphics[width=0.40\columnwidth]{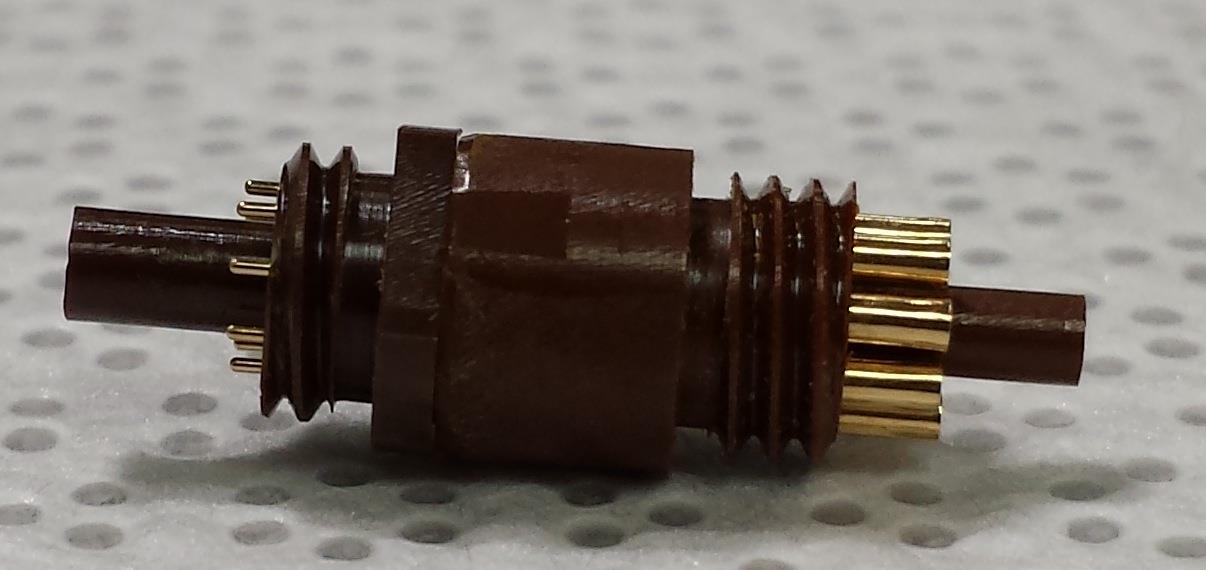}
    \includegraphics[width=0.60\columnwidth]{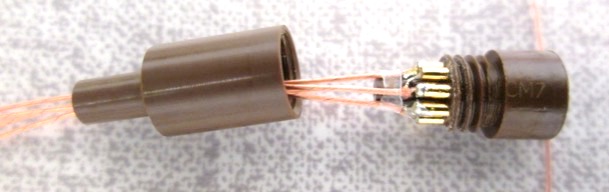}
    \includegraphics[width=\columnwidth]{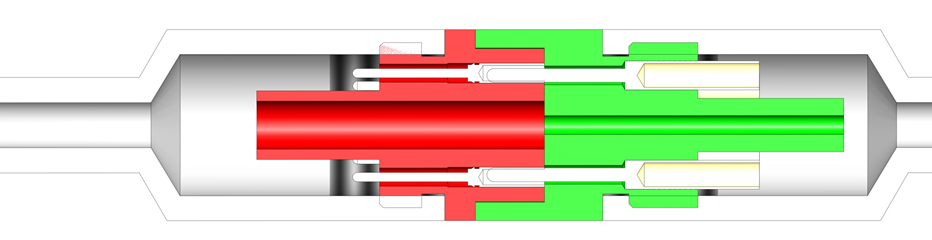}
    \includegraphics[width=0.34\columnwidth]{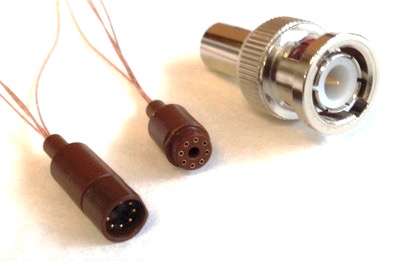}
    \includegraphics[width=0.66\columnwidth]{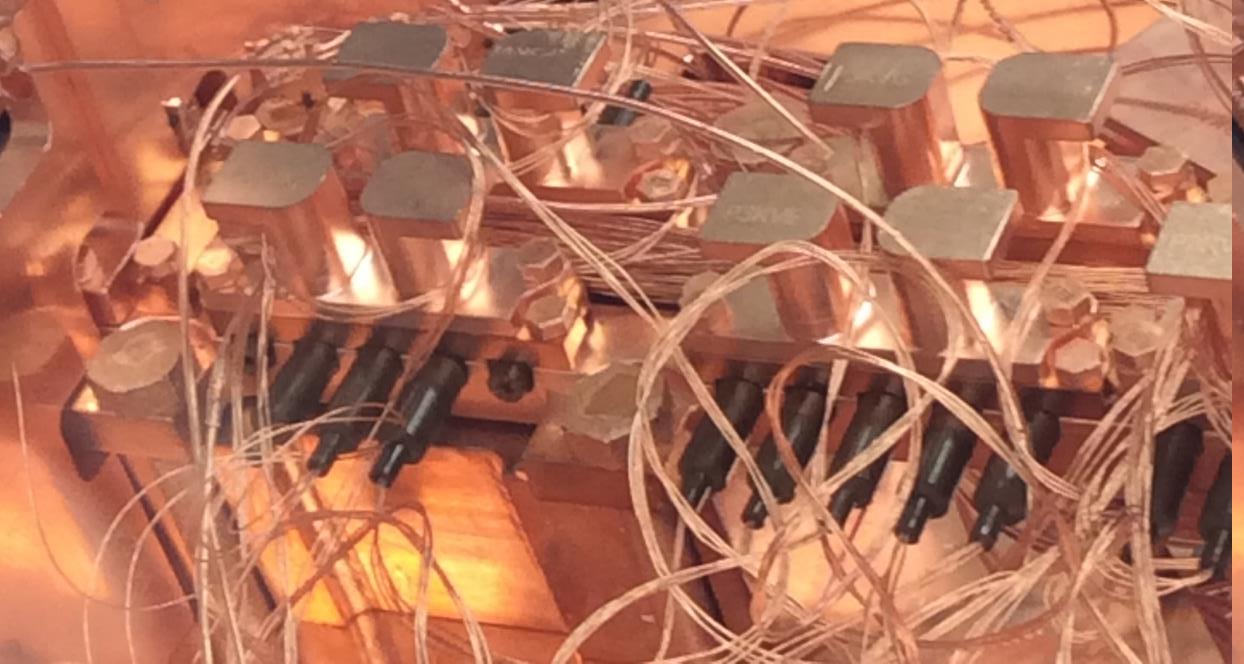}
    \caption{(Top left)  A photo of the custom-made connector without the cable
    housing.
    (Top right) A photo of the male end of the connector with the cable housing opened
    to show signal cables connected with tin-silver eutectic, and 
    strain relief provided by FEP shrink tubing.
    (Middle) Internal schematic of the signal connectors, showing the
    pin misalignment that provides the contact force (female connector in red,
    male connector in green).
    (Bottom left) A fully assembled male connector (left) and female connector
    (center), next to a BNC connector for scale.  When connected, the custom-made
    connector is 6.35~mm in diameter and 20.96~mm in length.
    (Bottom right) Several cables and connectors were installed above the cold plate.\label{fig:signal-connectors}}
\end{figure}

Above the cold plate, custom connectors~\cite{Busch:2017kxq} are needed to meet 
\textsc{Majorana Demonstrator}'s  background specifications. Electrical contact springs
in commercially available connectors typically contain beryllium-copper (BeCu), which has
unacceptably high radioactivity.  Contact force is necessary to maintain a
robust connection under temperature cycling.  A spring-free design is used, using
commercial Mill-Max gold-plated brass pins (8210 Receptacle with a
Standard Tail) produced in a special run without BeCu spring inserts, 
housed in a Vespel body custom-machined in the
\textsc{Majorana Demonstrator}'s underground machine shop.  The pins
and sockets are slightly misaligned radially, so that the contact force is provided by the forced
bending of the pin as it is inserted into the socket.  A low-background
tin-silver eutectic developed for the SNO experiment~\cite{SNOEutectic} is used
to solder the cables to the connector pins and sockets.  FEP shrink-tubing
provides strain relief for these connections (Figure~\ref{fig:signal-connectors}).

High-precision machining is required for fabricating these custom connectors to maintain a secure connection. But electrical discontinuity was
observed during thermal cycling in a subset of these connectors.  Additionally, some
connections in successfully bench-tested D-sub commercial connectors failed in the field,
reducing the number of available spare channels.  Possible explanations for
these failures include mishandling at installation, inadequate strain relief, and
deflection of the connector body that prevents robust contact in the central
region of the connector. In at least one instance, electrical shorts have been observed
between the signal cable ground shields and the cold plate, implying damage to the
signal cables during installation or handling. An upgrade that improves cable routing and connector design completed in early 2020.  These improvements will be reported in a future publication on post-upgrade performance of the \textsc{Majorana Demonstrator}.

\subsection{Internal high-voltage cabling}
The internal high-voltage cables also use the Axon'
PicoCoax\textsuperscript{\textregistered} design, but with thicker dielectric to
support high-voltage operation.  They consist of a 0.153-mm-diameter copper central conductor, a 0.311-mm-thick FEP inner dielectric (0.77-mm
outer diameter), a copper helical shield as ground (0.82-mm outer
diameter), and a 0.19-mm-thick outer jacket (1.2-mm outer diameter).  The resulting
cable is rated to 5~kV DC, and has an approximate linear mass density of 3~g/m.  

At the vacuum side of the feedthrough flange, each central conductor and ground
terminate in commercial BeCu D-sub sockets, housed within a
custom-machined 2-pin polyetheretherketone (PEEK) body. The resulting plug is push-connected to the
corresponding HV and ground pins on the feedthrough flange.

The connection between the HV cable and a detector's HV contact ring occurs
through contact with a spade-lug connector (the ``HV fork''), custom-machined out of
electroformed copper.  Close to this connector, the outer jacket and shielding
of the HV cable is stripped back and protected by FEP shrink tubing to ensure no contact with
ground. At the connector, the bare central conductor of the HV cable is woven
through two holes, and fixed in the second hole with a Vespel plug.

\begin{figure}
\begin{centering}
    \hspace{0.5cm}
    \includegraphics[width=0.97\linewidth]{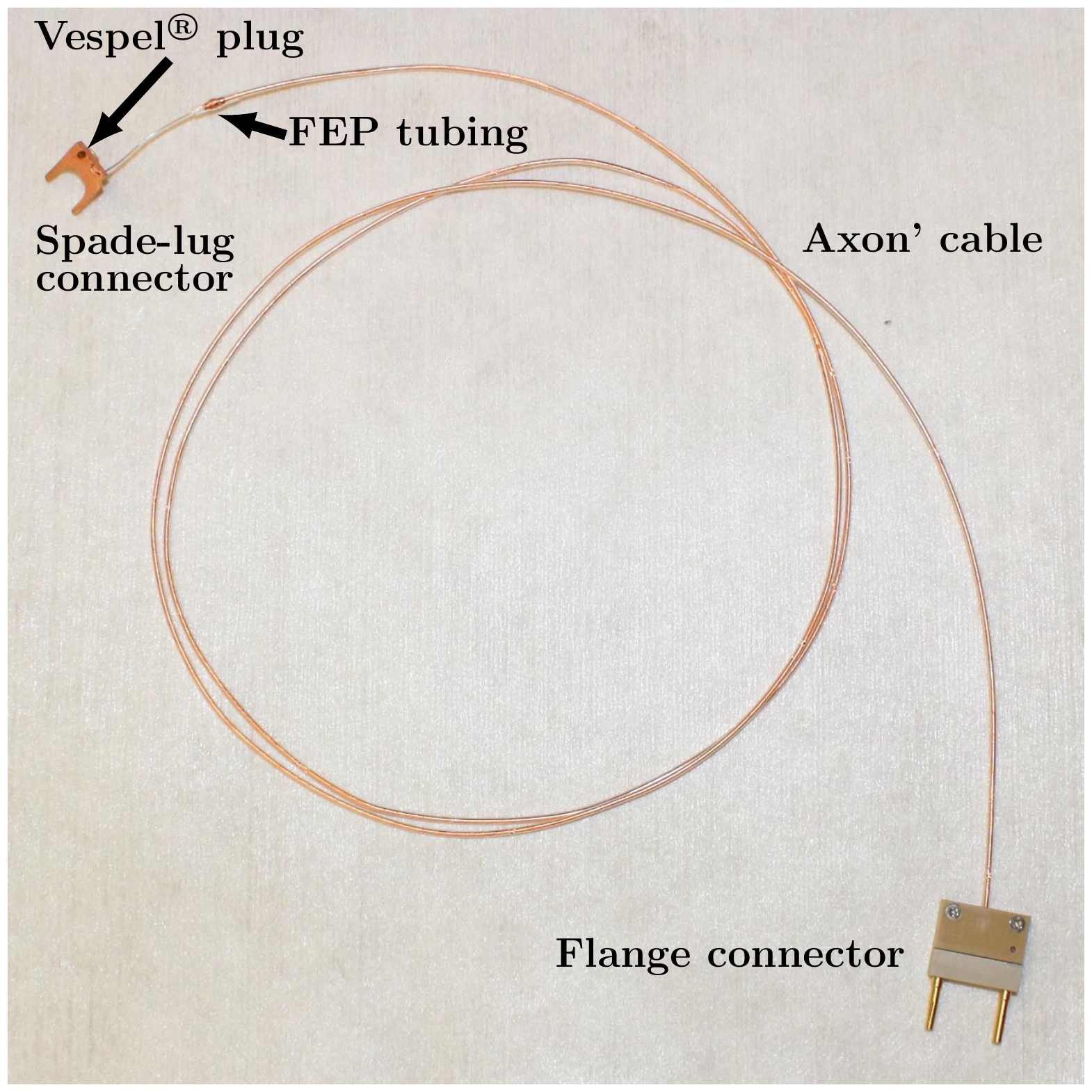}
    \caption{Picture of an HV cable with the HV fork connector on the top left, and the flange connector on the bottom right.}
    \label{fig_hv_cable_and_connector}
    \hspace{0.5cm}
\end{centering}
\end{figure}

The high-voltage cables, connectors, and feedthroughs were tested
to ensure that micro-discharges would not be induced
by the detector bias voltages~\cite{Abgrall:2016act}. All HV cables were
tested prior to installation.  Nonetheless, the 
high-voltage cables and connectors also have operational issues.

The most prevalent were the HV
breakdowns observed as significant discharges in the cables during
their initial operations.  To prevent damages to the electronics, these
detectors were fully or partially biased down while the cause of the breakdowns
was investigated.  Once it was determined that the breakdowns were occurring
between the HV cable central conductor and ground shield, 11 detectors were
brought online by not connecting the shield to ground. Additional tests on
spare cables revealed excessive compression during installation as the
likely cause of the deformation to produce this
problem.  Improvements in cable routing and the designs of in-vacuum HV connectors were realized in the upgrade.

\section{Back-end electronics}
\label{sec_back_end_electronics}
\begin{figure}[!htpb]
        \includegraphics[width=0.97\linewidth]{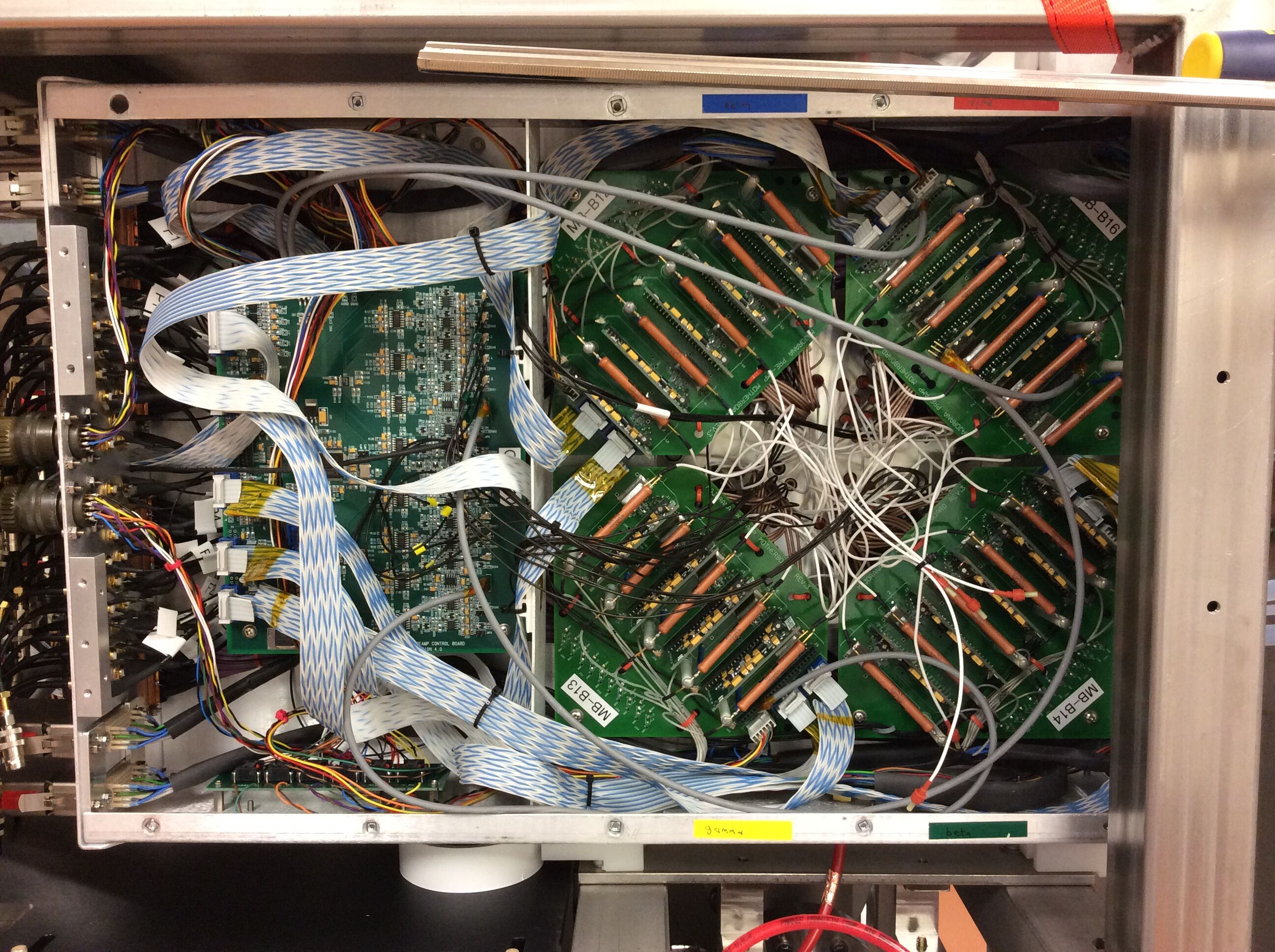}
        \caption{Picture of the contents inside an electronics box, showing a
        controller card on the left (the second controller card is beneath it), 
        and four motherboards on the right.  Connections to
        the vacuum feedthroughs are made in the central space between the
        motherboards.}
        \label{fig:EB}
\end{figure}

The back-end electronics consist of motherboards, preamplifier cards, and
controller cards. These are contained in electronics boxes mounted on
feedthrough flanges (two electronics boxes per detector module). Each
electronics box contains two controller cards, and
four motherboards that fit together with space to access the connections on the
air side of the feedthrough flange (Figure~\ref{fig:EB}).  A 
total of 40~W of heat is produced in a fully loaded
electronics box.  The electronics boxes are housed within the poly shield,
whose enclosed volume is actively cooled with a chilled water radiator and a
flow of boil-off nitrogen gas.

\subsection{Motherboard}
\label{subsec_MB}

\begin{figure}[!h]

\centering
    \includegraphics[trim=0 10cm 25cm 0, clip, width=0.50\textwidth]{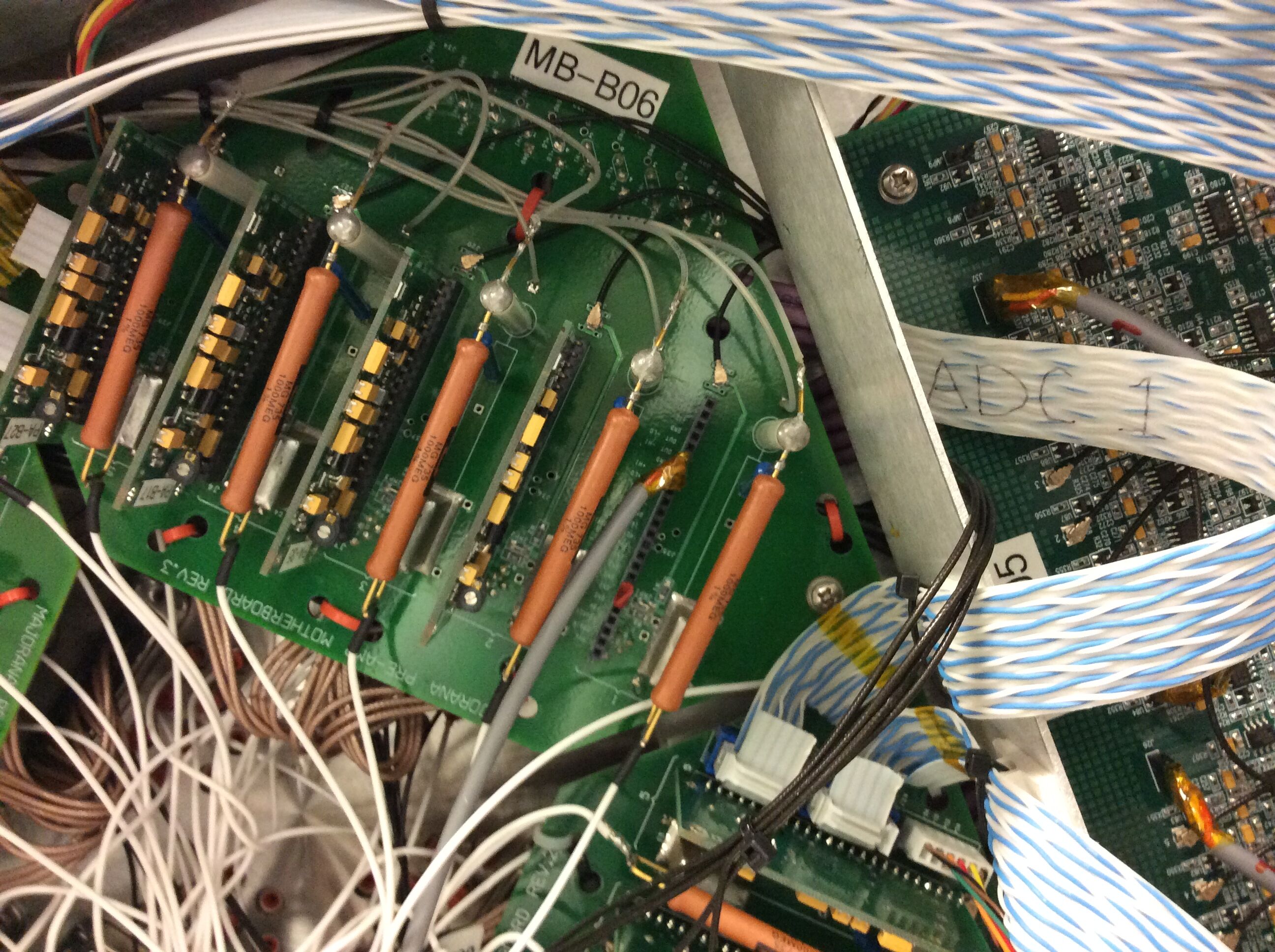}
    \caption{Picture of a motherboard installed in the electronics
    box. Preamplifier cards are installed in the left four slots, and
    the 1~G$\Omega$ resistors of the 5 HV filters are visible.\label{fig:MB}}
\end{figure}

Each string of detectors is served by one motherboard (Figure~\ref{fig:MB}).
The motherboard has connections for up to five preamplifiers, where the gain loop including the LMFE is closed.  
The preamplifier outputs are forwarded
to the digitizer card.  The mapping of strings to
motherboards was chosen to utilize working connectors and allow for an equal
distribution of spare channels throughout the electronics box.

The output of the high-voltage power supplies is also routed through the motherboard, 
which contains five high-voltage 
RC filters, each  consisting of a Caddock MG735 1-G$\Omega$ 1\% resistor
and a custom Novacap 8.2nF 7000V capacitor.  The HV filters also have a
1-G$\Omega$ 5\% bleeder resistor (Ohmite MOX1125J-1000ME-ND) to ground for safety.  
For testing and calibration, charges can be injected into the detector by bypassing the HV filter via a
voltage divider.

The low-voltage power supply outputs also route through the motherboard, which
distributes power to the preamplifiers and controller cards. The power
provided to the JFET on each LMFE is set by the controller cards, and routed
to them through the preamplifiers via the motherboard
(Section~\ref{subsec_CC}).  

The motherboard is a 4-layer board, utilizing separate ground
planes for the preamplifiers and the high-voltage filters. 
Communication between the controller card and the digitizer is also routed through the
motherboard.

\subsection{Preamplifier card}
\label{subsec_PA}

\begin{sidewaysfigure}
\begin{centering}
    \hspace{0.5cm}
    \includegraphics[width=0.97\linewidth]{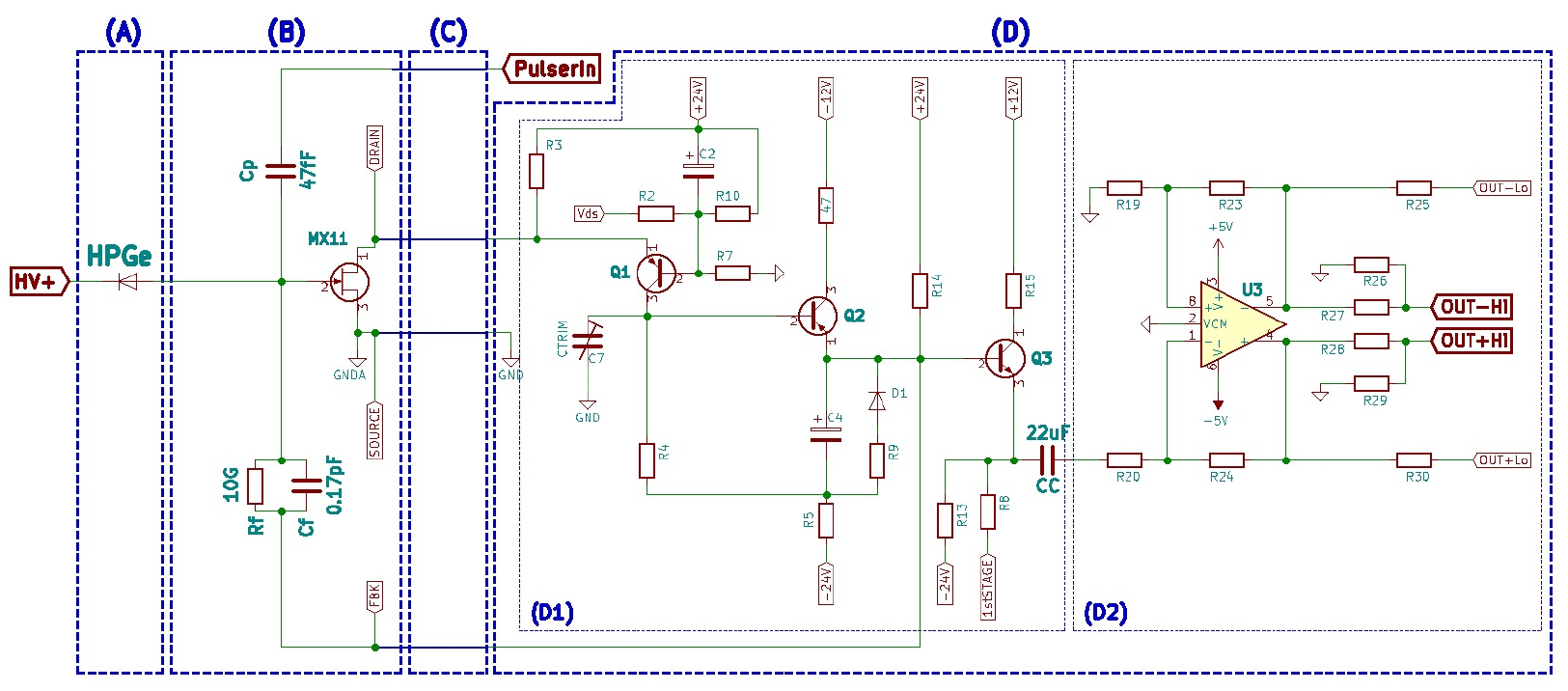}
    \caption{A simplified circuit diagram showing (A) the High-Purity Germanium (``HPGe") Detector with (B) the low-mass front-end (LMFE) electronics connected with (C) a 2.15-m pico-coaxial cable to (D) the preamplifier card. The detector, front-end board, and cable are both cold and in vacuum. The preamplifier card is located in the electronics box at room temperature and  pressure. The LMFE's amorphous germanium feedback resistor (``Rf") is in parallel with a trace capacitance (``Cf") of 0.17-pF. The trimmer capacitor (``CTRIM") is adjustable between 1 to 20~pF, and the coupling capacitor (``CC") is 22~$\mu$F. The preamplifier is divided in two sections: D1 and D2. D1, when connected to the LMFE makes the charge amplification circuit, while D2 is the differential buffer that outputs the amplified charge signal.}
    \label{fig_lmfe_and_preamp}
    \hspace{0.5cm}
\end{centering}
\end{sidewaysfigure}

To reduce the number of active components (hence background radioactivity) near
the detectors, the gain loop for detector signal amplification is spatially separated
(Figure~\ref{fig_lmfe_and_preamp}) between the warm preamplifier card on
a motherboard outside the detector shield 
and the cold LMFE board (Section~\ref{sec_front_end_electronics}).  Together they form a charge-sensitive preamplifier.

The preamplifier uses a folded-cascode design based on the Goulding amplifier~\cite{Goulding}.  
It differs from the Goulding amplifier in its choice of power supply lines
($\pm12$~V and $\pm24$~V DC) and in its implementation of a capacitively
coupled second amplification stage using operational amplifiers.
The output of the first amplification stage is available for diagnostic
purposes (e.g. measuring detector leakage current and signal rise-times), 
and is read out by an ADC in the controller card (Section~\ref{subsec_CC}).

The second amplification stage
provides differential outputs for two separate gains.
The dynamic range is $0$~to~$\sim3.5$~MeV for the high-gain output, and
$0$~to~$\sim10$~MeV for the low-gain output. No
pole-zero correction is performed by the preamplifier, opting instead for
post-processing of the digitized pulse. 

During commissioning, pulses were applied to each front-end's pulser line, 
and a 1-20~pF trimmer capacitor on each preamplifier was adjusted to remove
ringing in the output waveform. 

With careful choice of front-end components and a short cable connection, 
the preamplifier is capable of a sub-10-ns rise time. With the LMFEs and
full-length cables used in the \textsc{Majorana Demonstrator}, and the
adjustable capacitor on the preamplifier card tuned to
prevent ringing, rise times of 120-200~ns were observed at the output of the 
first stage of the preamplifier card during testing.
  
\subsection{Controller card}
\label{subsec_CC}
The controller cards house the pulser system electronics, as well
as circuitries for controlling and monitoring the readout electronics.  Each controller card performs these functions for two strings of
detectors.

Each controller card contains one Xilinx XC3S200A-4VQG100C Field Programmable Gate Array (FPGA), which
is the interface between the optically-coupled SPI bus to the digitizer and the data conversion chips
on the controller card.  All settings and measurements communicated between the
digitizer and controller cards are available in the ORCA slow-control system.

Two Analog Devices AD7327BRUZ analog-to-digital
converter (ADCs) provide 16 analog inputs.  
Four inputs monitor the preamplifier supply voltages, and two
inputs are used to monitor the on-chip temperature.  The remaining ten inputs
monitor the first-stage output of the preamplifiers.

The controller card also contains two Linear Technology LTC2600CGN 16-bit
digital-to-analog converter (DAC) chips, whose outputs set the drain-to-source
voltage (V$_{\mathrm{DS}}$) of the JFET on each of the front-end boards individually.  Adjusting  
V$_{\mathrm{DS}}$ fine-tunes the JFET performance by varying
its power.

Each controller card contains two pulse generator systems, each including one
LTC2600CGN~DAC, 
four Analog Devices AD8182AR 10-ns switching multiplexers, and eight operational-amplifier-based
attenuators, providing a combined total of 16 pulse-generator outputs.
These pulse generators are collectively referred to as \emph{the pulser system}. 
The amplitude, high/low time, number of pulses, and attenuation are
user-specified in the ORCA interface.  During normal operation, the pulser period
set on each controller card is typically $\sim8.5$~s.  Pulser events are used
primarily for livetime monitoring, to confirm detector operation in the absence
of physics events (due to the low event rate in each detector).  They are also
used to monitor gain stability and trigger efficiency, and have been used for the 
validation of digitizer linearity~\cite{MAJORANA:2020llj}. 

\section{Digitizer and power supplies}
\label{sec_power_supplies_and_digitizer}

\subsection{Digitizer}
\label{subsec_digitizer}
The amplified differential signal is digitized by the
GRETINA~\cite{Gretina} digitizer card. The card has 10 differential input
ports; thus, one card can serve one string of
detectors (five detectors with two gains each). Each card also
has an auxiliary digital I/O port, which is used to communicate with a controller card (two
strings) over the SPI bus.

Incoming waveforms are digitized at a rate of 100 MSample/s by 14-bit 105-MHz flash ADCs, 
providing a two's complement signed
integer for each sample. Each card has an on-board pre-ADC leading-edge
discriminator (LED) trigger, but the \textsc{Majorana Demonstrator} uses a
trapezoidal trigger for practically all data taking.  This  trigger uses the
value of the onboard-calculated trapezoidal filter as a trigger parameter.
The time synchronization between all digitizer cards is performed by
a `trigger card' hosted in the same VME crate that distributes clock
signals to all digitizers. Different onboard features can be set, such as
multisampling,
pole-zero (PZ) correction, and trapezoidal filtering. Energy is re-calculated
during data processing, along with corrections such as PZ and ADC
non-linearities~\cite{Alvis:2019sil,MAJORANA:2020llj}.  The information produced by the digitizer card for
a single event/trigger consists of a timestamp, an uncalibrated energy
estimate, and a $\sim2000$-sample waveform.

With the exception of a three-month period in 2016, all data taken prior to May
2017 used 10-ns time samples, giving a waveform of $\sim20~\mu$s in length.
During that period and after May 2017, the data has been taken in
multi-sampling mode.  In this mode, the first $\sim14$~$\mu$s of the waveform is
digitized using 10-ns time samples and the samples after that are 40-ns long, giving
such $\sim2000$~sample waveform a length of $\sim38$~$\mu$s.  Multi-sampling retains full detail in the
rising edge of the peak needed for multisite-event
rejection~\cite{Alvis:2019dzt}, while allowing digitizing more of the falling
part of the waveform to improve alpha-background rejection~\cite{Gruszko:2016hfh,Majorana:2020xvk}.

\subsection{High-voltage power supplies}
\label{subsec_HV}

The high voltage to bias the detectors is provided by ISEG EHS8260p\_105 modules hosted in a WIENER MPOD 
crate.  These power supplies have low ripples ($<20$~mV), which is needed for good energy
resolution and pulse shape analysis performance.
The modules can deliver up to $+6$~kV on 8 channels with return
lines isolated (i.e. no common ground). The HV is controlled through
customized scripts (e.g. ramp-up, ramp-down) in
the ORCA interface. Each detector is limited in ORCA to never exceed its target
operating voltage.  In no situation can a detector voltage exceed 5~kV even if
the voltage target was improperly set. The high-precision readout
mode was enabled in the SNMP Ethernet communication with the MPOD
controller card to closely monitor the current delivered by the
supplies.  If this current exceeds an ORCA-preset maximum threshold of
20~$\mu$A, the HV is shut down to
protect the detectors.  The safety-loop feature of the EHS8260p\_105 modules was also
hardware-enabled for integration in the interlock system described
in the next section.

\subsection{Low-voltage power supplies}
\label{subsec_LV}
A custom-designed low-noise power supply provides power to the controller cards,
preamplifier cards, and the JFET on each front-end. 
The low-voltage power supply module provides six independent voltage outputs
from regulated linear supplies, each with isolated returns. Four rails
($\pm$24 and $\pm$12~V) are distributed to the preamplifiers, while
the last two ($\pm$7.5~V) and an unregulated rail ($+$12~V) are
used for the controller card. As described in
Figure~\ref{fig_electronics_general_layout}, one LV module provides
power for two strings of detectors and their associated controller
card.

For interlocking with the HV modules, a relay-based logic was
implemented to divert a small fraction of current from the unregulated
12~V supply if all of the other supplies provide the expected
outputs. This current flows through the safety-loop
circuit of the ISEG HV modules that power the same strings of
detectors. If the LV is off or turned off, no current flows through this loop
and the HV supplies will be biased down
to protect the front-end electronics.

\section{General grounding scheme}
\label{sec_general_grounding_scheme}
In the \textsc{Majorana Demonstrator} underground laboratory, the ``general ground'' is a copper stake
driven into the floor of the electrical utility room, where the electrical
breakers are located.  The experiment is
in a separate room (``the detector room'').  The AC power lines from the
electrical breakers to the detector room provide one connection to the general
ground.  There is also a ground plate mounted low on the wall in the detector
room with a dedicated low-resistance ground cable to the general ground.

\begin{figure}[!htb]

    \includegraphics[width=\columnwidth]{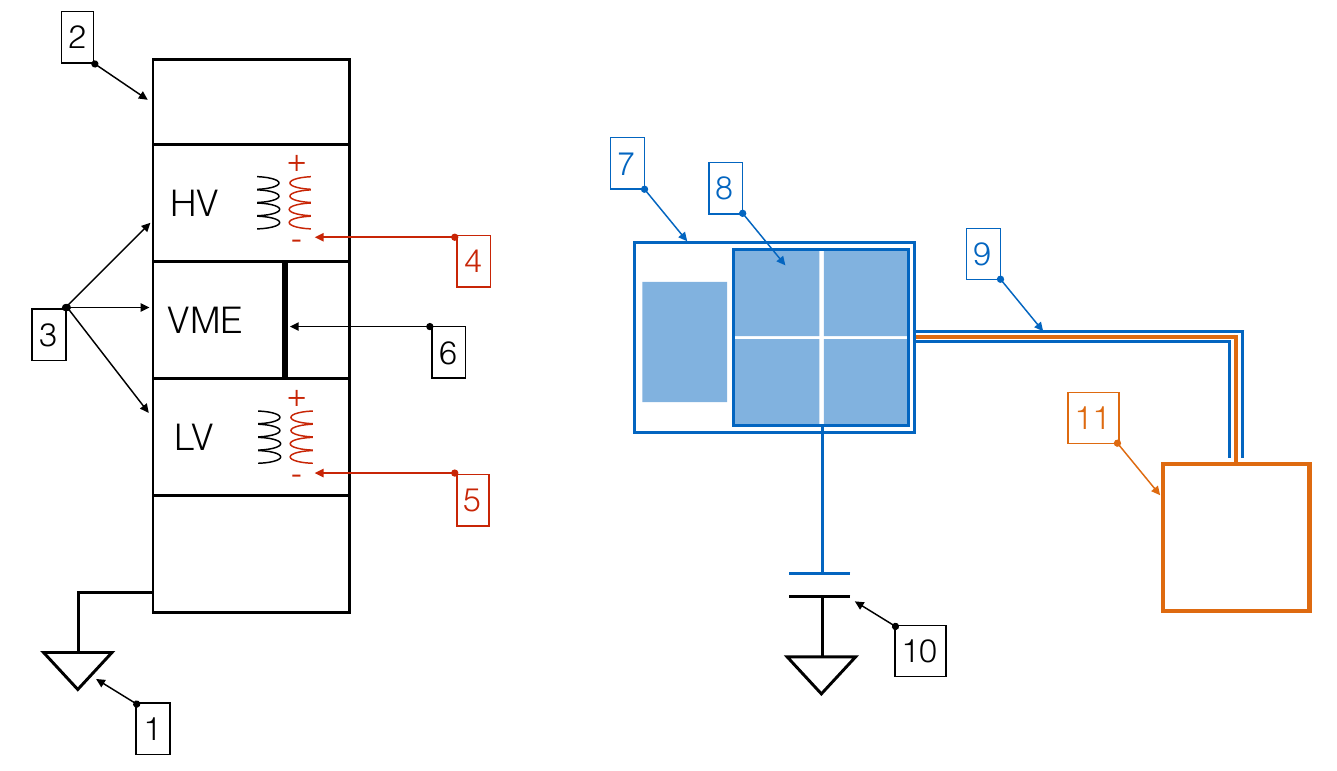}
    \caption{Grounding scheme of the \textsc{Majorana Demonstrator}. The ``equipment racks''
    conductive body consists of items 2, 3, and 6.  The ``exterior-of-module''
    conductive body consists of items 7, 8, and 9.  The ``interior-of-module''
    conductive body consists of item 11.  Items 4 and 5 are grounded to the
    ``exterior-of-module'' conductive body.
    Each conductive body's path to ground is described in
    Section~\ref{sec_general_grounding_scheme}.  The labeled items are: (1) the general ground in
    the floor of the breaker room; (2) electronics racks; (3) crate chassis; (4) HV
    module returns; (5) LV module returns; (6) digitizer card ground planes; (7)
    electronics box chassis and exterior of the cryostat and crossarm;
    (8) ground planes of boards in electronics box; (9) shield of internal
    cables; (10) capacitive coupling to ground via metallic support structures; and
    (11) thermosyphon and internal structural copper parts. \label{fig:grounding}}
\end{figure}

In the general grounding scheme (Figure~\ref{fig:grounding}), electronics components 
are grouped by their location into three isolated conductive bodies in the
detector room: the equipment racks, the
exterior of the module (components outside the vacuum), and the interior of the
module (the thermosyphon and internal structural copper parts.)  The connection
between these bodies and ground was carefully planned and executed in order to
minimize electronics noise from ground loops.

The three equipment racks are conductive bodies: one for each of
the two modules housing the power supplies (HV and LV) and DAQ electronics, and
one for slow controls and monitoring.  The computers in each rack are powered by the outlets in the 
iBootBars (Dataprobe iBootBar N15). Each outlet can be remotely
controlled for rebooting of equipment and computers.  The iBootBars connect
to the general ground via the AC power lines.  Each equipment rack is grounded via its connection to 
the iBootBars.  The HV and digitizer
modules are therefore grounded through their crates to the rack, and the LV
power supplies are grounded through their cases to the rack.  The returns of the HV
modules and LV power supplies are all
isolated from the conductive body of the equipment racks, instead of being part of the
exterior-of-module conductive body.

The ``exterior-of-module'' conductive body contains the electronics outside each
module, i.e. the electronics boxes (controller cards, motherboards, and
preamplifier cards) as well as the wall of each module's cryostat, and the
exterior of the crossarm.  These bodies are separated by glass isolators and plastic
supports from the surrounding metallic support structures, which
have their own paths to ground. The ground planes of the controller cards,
motherboards, and preamplifier cards are all grounded to the electronics-box
case through metallic standoffs.  The power cables from the LV power supplies
to the electronics boxes are grounded to the respective box.  
The cables from the digitizers and the HV power supplies
to the electronics box are grounded at both ends, providing the connection of
the exterior-of-module conductive body to general ground through the
equipment racks.

The ``interior-of-module'' conductive body consists of the internal 
structural copper parts (thermosyphon, cold plate, and string
assemblies).  At each of the module's two feedthrough flanges, a spare HV cable
is plugged into an unused HV channel ground wire and bolted to the cold plate.
This grounds the interior-of-module to the exterior-of-module
conductive body.  A glass isolator between the copper thermosyphon and the rest of
the cooling system ensures that these cables are the only ground path for the
interior-of-module conductive body.

\section{Spectroscopic performance}
\label{sec:spec_perf}
\begin{figure}[!htb]

    \includegraphics[trim=0 0 0 2.5cm, clip, width=\columnwidth]{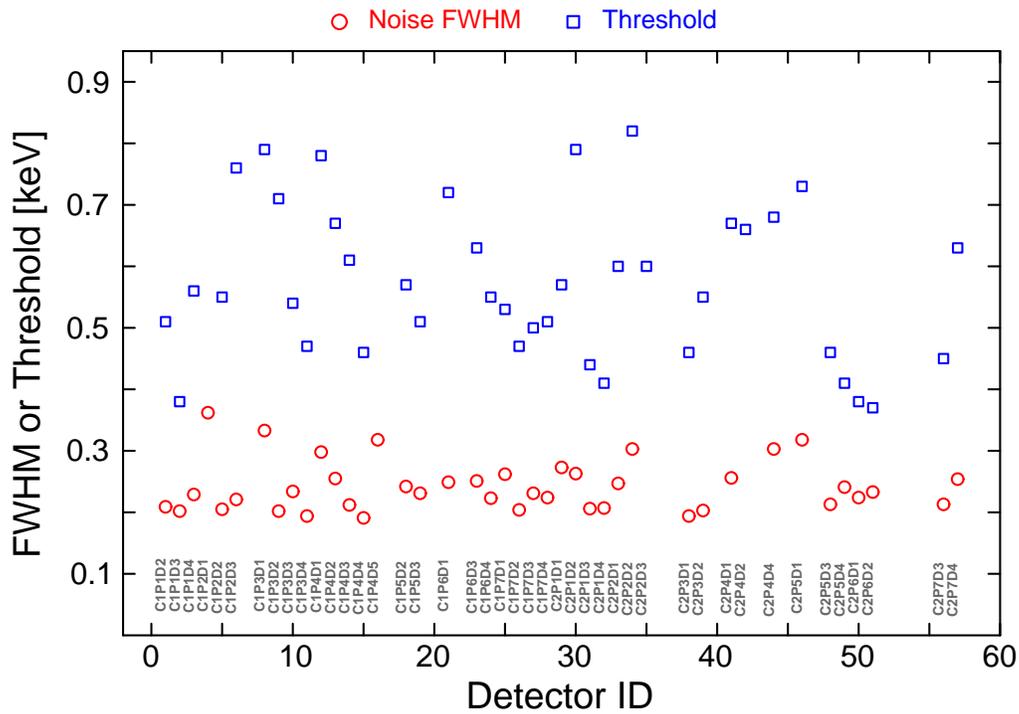}
    \caption{The measured FWHM (circles) and the trigger threshold
    (squares) for each detector measured during a typical physics run.  The average noise 
    is typically less than 250~eV and the average triggering threshold is normally less
    than 700~eV.
    \label{fig:noise-thresh}}
\end{figure}

The spectroscopic performance of the full production readout chain is summarized in this section. In Figure~\ref{fig:noise-thresh},
the measured noise (FWHM) and threshold of each detector in a typical physics run
are shown. The noise is calculated using the RMS of the signal baseline samples.  The threshold is taken from the value loaded into the GRETINA digitizer, corrected for a small offset in the on-board threshold trapezoid value that can change (and is determined independently) after every initialization of the digitizer board.  Both are converted to energy values based on a rough energy-scale calibration (which is accurate to within 5\%).
n
The average noise level of $\le250$~eV is crucial in delivering good energy
resolution.  The sub-keV thresholds enable low-energy searches 
for physics beyond the Standard Model.  The \textsc{Majorana Demonstrator} is
the only neutrinoless double-beta decay experiment that triggers at such 
low energies to engage in such searches.

\begin{figure}[!htb]
    \includegraphics[width=\columnwidth]{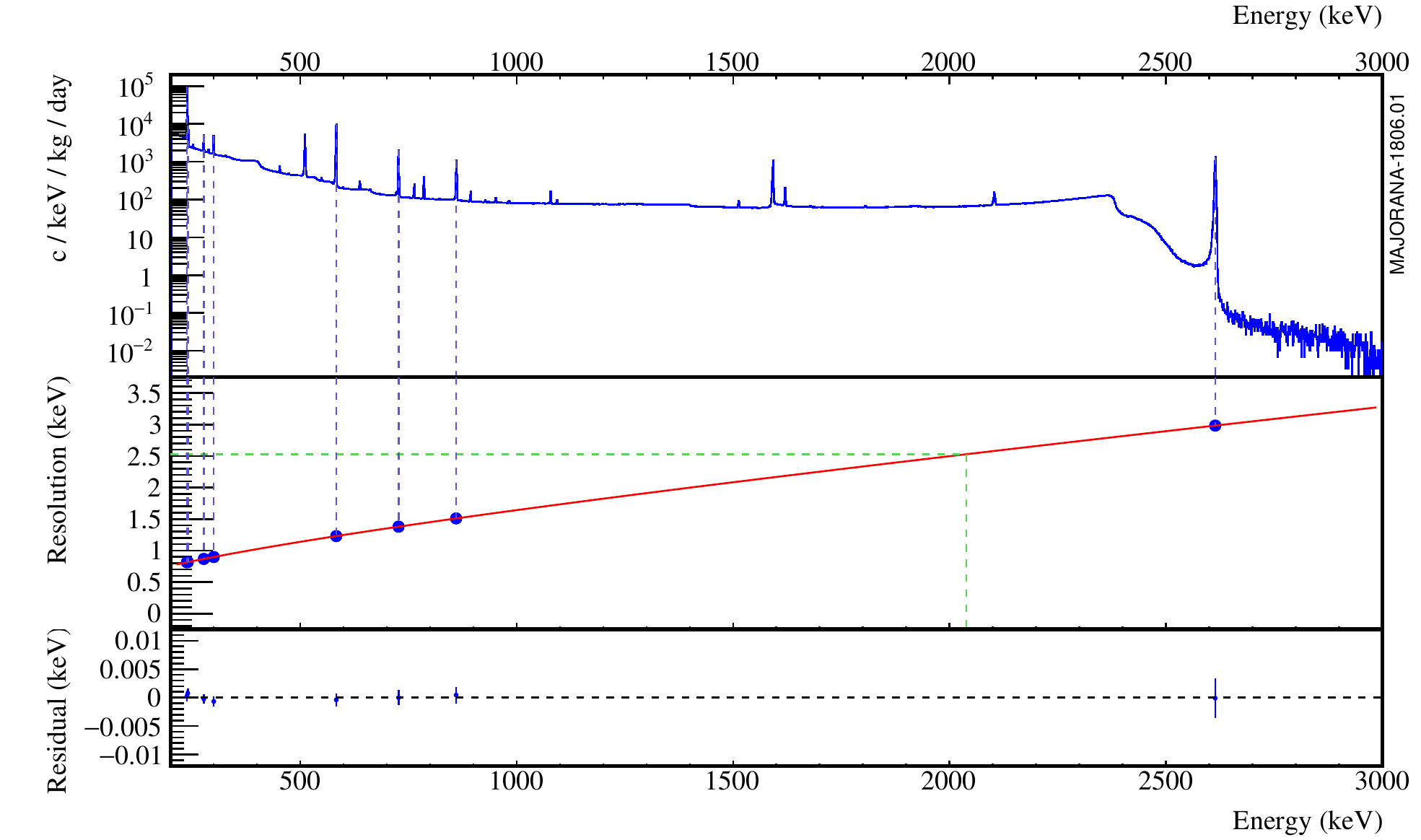}
    \caption{An energy spectrum of Th calibration data (top) with the
    fitted FWHM energy resolution (middle) and the fit residuals (bottom).  The energy resolutions of the major peaks are
    shown along with the
    energy resolution at the 2039~keV Q-value (dashed green lines).
    \label{fig:spectrum-eres}}
\end{figure}

\begin{figure}[!htb]
    \includegraphics[width=\columnwidth]{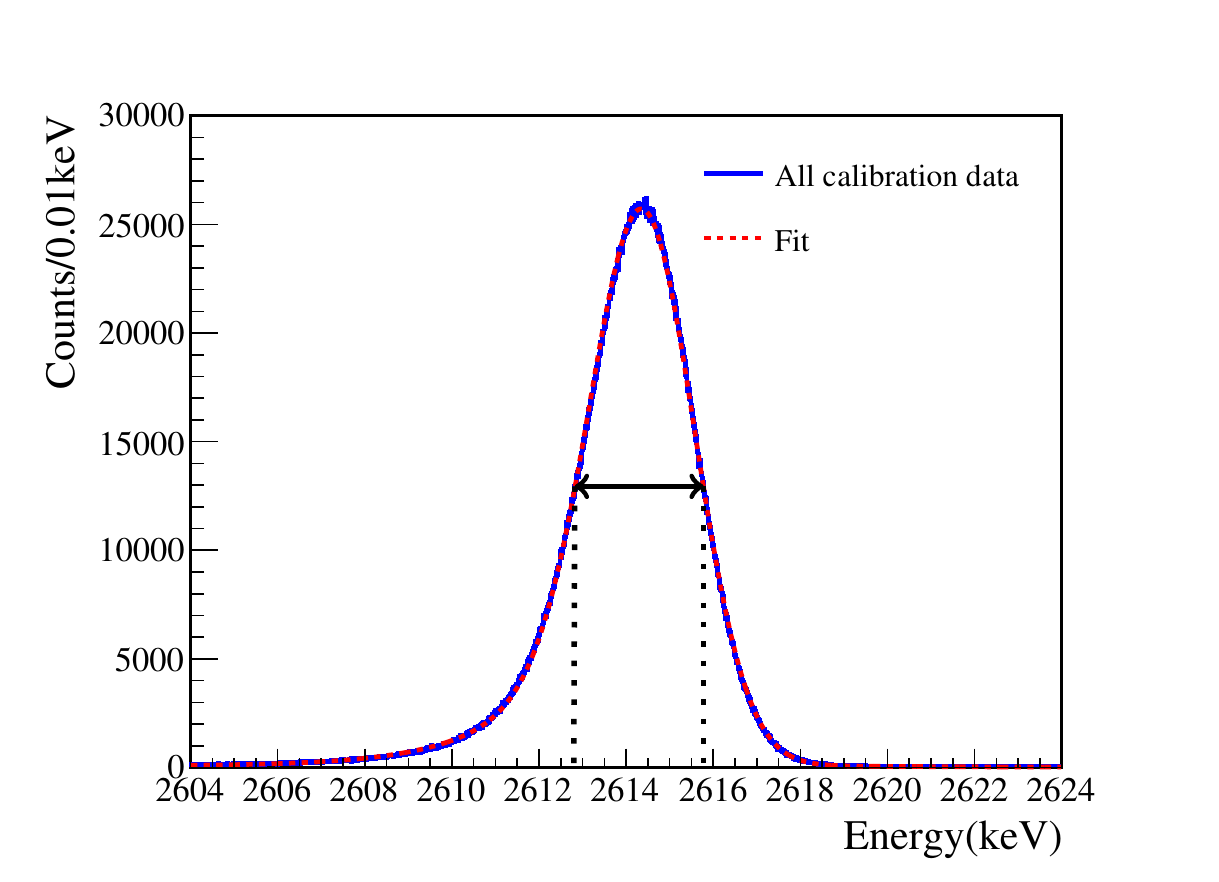}
    \caption{The $^{208}$Tl peak (2614.5~keV) from Th calibration data and its fit.  An  energy resolution of 3.0~keV (FWHM) is measured at this energy.
    \label{fig:peak-eres}}
\end{figure}

In Figure~\ref{fig:spectrum-eres} the full energy spectrum from a Th-source
calibration is shown along with the measured energy resolution as a function
of energy.  The energy resolution of 0.12\% at the Q-value of 2039~keV~\cite{Aalseth:2017btx} is the
best energy resolution of any operating neutrinoless double-beta decay
experiment.  A closer look at the $^{208}$Tl calibration peak-model fit to the data is shown
in Figure~\ref{fig:peak-eres}.

\section{Conclusions}
The commissioning of the \textsc{Majorana Demonstrator} began in June 2015,
followed by data production with the full detector array in August 2016.  Custom-designed
front-end and back-end electronics, low-voltage power supplies, and cables and connectors, were required to
achieve the necessary low-noise and low-background performance.
Pre-existing solutions were used for signal digitization and high-voltage power
supplies.
The electronic noise was further minimized by a grounding scheme carefully planned and implemented
to avoid ground loops. This readout electronics system design has led to
a neutrinoless double-beta decay program with world-leading energy resolution 
and low-energy searches for additional physics beyond the Standard Model.

\section{Acknowledgments}
This material is based upon work supported by the U.S.~Department of Energy,
Office of Science, Office of Nuclear Physics under contract / award numbers
DE-AC02-05CH11231,  \\ DE-AC05-00OR22725, DE-AC05-76RL0130, DE-FG02-97ER41020,
DE-FG02-97ER41033, \\ DE-FG02-97ER41041, DE-SC0012612, DE-SC0014445, DE-SC0018060, and LANLE9BW. We acknowledge support from the Particle Astrophysics Program and Nuclear Physics Program of the National Science Foundation through grant numbers MRI-0923142, PHY-1003399, PHY-1102292, PHY-1206314, PHY-1614611, PHY-1812409, and PHY-1812356. We gratefully acknowledge the support of the U.S.~Department of Energy through the LANL/LDRD Program, and through the LBNL/LDRD and PNNL/LDRD Programs for this work. We acknowledge support from the Russian Foundation for Basic Research, grant No.~15-02-02919. We acknowledge the support of the Natural Sciences and Engineering Research Council of Canada, funding reference number SAPIN-2017-00023, and from the Canada Foundation for Innovation John R.~Evans Leaders Fund.  This research used resources provided by the Oak Ridge Leadership Computing Facility at Oak Ridge National Laboratory and by the National Energy Research Scientific Computing Center, a U.S.~Department of Energy Office of Science User Facility. We thank our hosts and colleagues at the Sanford Underground Research Facility for their support.

\bibliographystyle{JHEP}
\bibliography{electronics-paper}

\end{document}